\documentclass{pasa}%

\usepackage{graphicx}
\usepackage{longtable}
\usepackage{ltablex}
\usepackage{comment}
\usepackage{threeparttable}
\usepackage{multirow}
\usepackage{rotating}
\usepackage[export]{adjustbox}
\newcommand{\arcsec}{^{\prime \prime}}
\newcommand{\arcmin}{^{\prime}}
\newcommand{\farcm}{\mbox{\ensuremath{.\mkern-4mu^\prime}}}
\newcommand{\farcs}{\mbox{\ensuremath{.\!\!^{\prime\prime}}}}
\newcommand{\fdg}{\mbox{\ensuremath{.\!\!^\circ}}}
\newcommand{\arcdeg}{\ensuremath{^{\circ}}}
\def\@mcsc{ptmrc}

\title[SkyMapper Southern Survey DR2]{SkyMapper Southern Survey: Second Data Release (DR2)}
\author[Onken et al.]{Christopher A. Onken$^{1,2}$, Christian Wolf$^{1,2,3}$, 
Michael S. Bessell$^1$, Seo-Won Chang$^{1,2,3}$, Gary S. Da~Costa$^1$, Lance C. Luvaul$^1$, Dougal Mackey$^1$, 
Brian P. Schmidt$^{1,2}$, Li Shao$^{1,4}$
\affil{$^1$Research School of Astronomy and Astrophysics, Australian National University, Canberra, ACT 2611, Australia}
\affil{$^2$Australian Research Council (ARC) Centre of Excellence for All-sky Astrophysics (CAASTRO)}
\affil{$^3$Australian Research Council (ARC) Centre of Excellence for Gravitational Wave Discovery (OzGrav)}
\affil{$^4$Kavli Institute for Astronomy and Astrophysics, Peking University, 5 Yiheyuan Road, Haidian District, Beijing 100871, P.~R.~China}
}%

\jid{PASA}
\doi{10.1017/pas.\the\year.xxx}
\jyear{\the\year}

\usepackage{aas_macros}
\usepackage{hyperref} 
\hypersetup{colorlinks,citecolor=blue,linkcolor=blue,urlcolor=blue}

\begin{document}

\begin{frontmatter}
\maketitle

\begin{abstract}
We present the second data release (DR2) of the SkyMapper Southern Survey, a hemispheric survey carried out with the SkyMapper Telescope at Siding Spring Observatory in Australia, using six optical filters: $u, v, g, r, i, z$. DR2 is the first release to go beyond the $\sim 18$~mag (10$\sigma$) limit of the Shallow Survey released in DR1, and includes portions of the sky at full survey depth that reach $>21$~mag in $g$ and $r$ filters. The DR2 photometry has a precision as measured by internal reproducibility of 1\% in $u$ and $v$, and 0.7\% in $griz$. More than 21\,000~$\deg^2$ have data in some filters (at either Shallow or Main Survey depth) and over 7\,000~$\deg^2$ have deep Main Survey coverage in all six filters. Finally, about 18\,000~$\deg^2$ have Main Survey data in $i$ and $z$ filters, albeit not yet at full depth. The release contains over 120\,000 images, as well as catalogues with over 500 million unique astrophysical objects and nearly 5 billion individual detections. It also contains cross-matches with a range of external catalogues such as {\it Gaia} DR2, Pan-STARRS1 DR1, {\it GALEX} GUVcat, 2MASS, and AllWISE, as well as spectroscopic surveys such as 2MRS, GALAH, 6dFGS, and 2dFLenS.
\end{abstract}

\end{frontmatter}

\section{INTRODUCTION}

The SkyMapper Southern Survey (SMSS) is obtaining a digital image of the entire Southern hemisphere of the sky. It is designed to reach a depth of 20 to 22 mag in six optical filters and achieve near-arcsecond-level spatial resolution. The survey started in 2014 with an emphasis on the short-exposure Shallow Survey. Images from the Shallow Survey reach 10$\sigma$ limits for point sources around 18~mag, and instantaneous six-filter photometry for over 280 million stars, galaxies and quasars were published in a first Data Release \citep[DR1;][]{Wolf18a}. 

DR1 has already facilitated the discovery of the most luminous quasar currently known \citep{Wolf18b}, while the discovery of the most iron-poor star currently known \citep{Keller14} was based on commissioning data from SkyMapper. On both topics, further work continues: a paper describing a new large sample of extremely metal-poor stars based on DR1 photometry and low-resolution spectroscopy with the Australian National University 2.3m telescope is nearing completion (Da Costa et al., in preparation), as is a paper discussing the results from high-dispersion spectroscopic follow-up of the most metal-poor star in this sample \citep{Nordlander19}; similarly, the growing sample of confirmed ultraluminous quasars at $z>4$ has been compiled \citep{Wolf20}.

DR1 has also helped with discovering extremely metal-poor stars in the Tucana II Dwarf Galaxy \citep{Chiti18}, with studies of the most metal-poor Galactic globular cluster \citep{Simpson18}, and with characterising the lowest-mass ultra metal-poor star known \citep{Schlaufman18}. Many other scientific endeavours in the Southern skies are underway using DR1 data.

This paper now presents the second data release, DR2, which adds, for the first time, long exposures from the SkyMapper Main Survey. The 100~sec exposures of the Main Survey provide individually a $1-3$~mag gain in point source depth and surface brightness sensitivity. By the time the survey finishes, the Southern sky should be imaged to 5$\sigma$-limits of (20,21,22,22,21,20)~mag in the filters ($u,v,g,r,i,z$), respectively. All SkyMapper magnitudes are in the AB system \citep{Oke83}.

The leap in depth relative to DR1 extends the reach of Galactic archaeology studies in our own Milky Way, such as studies of Blue Horizontal Branch stars, which were limited in distance by the previous shallow DR1 \citep{Wan18}. Given the enhanced surface brightness sensitivity, DR2 now enables a broad range of work on galaxies, where the previously released Shallow Survey data of DR1 mostly supported studies of the Milky Way and bright quasars. The first example is by \citet{Wolf19} who present colour maps of well-resolved galaxies at low redshift, and discuss how SkyMapper filters help to trace spatio-temporal variations of the star-formation rate in galaxies.

\begin{figure*}
\begin{center}
\includegraphics[width=\textwidth]{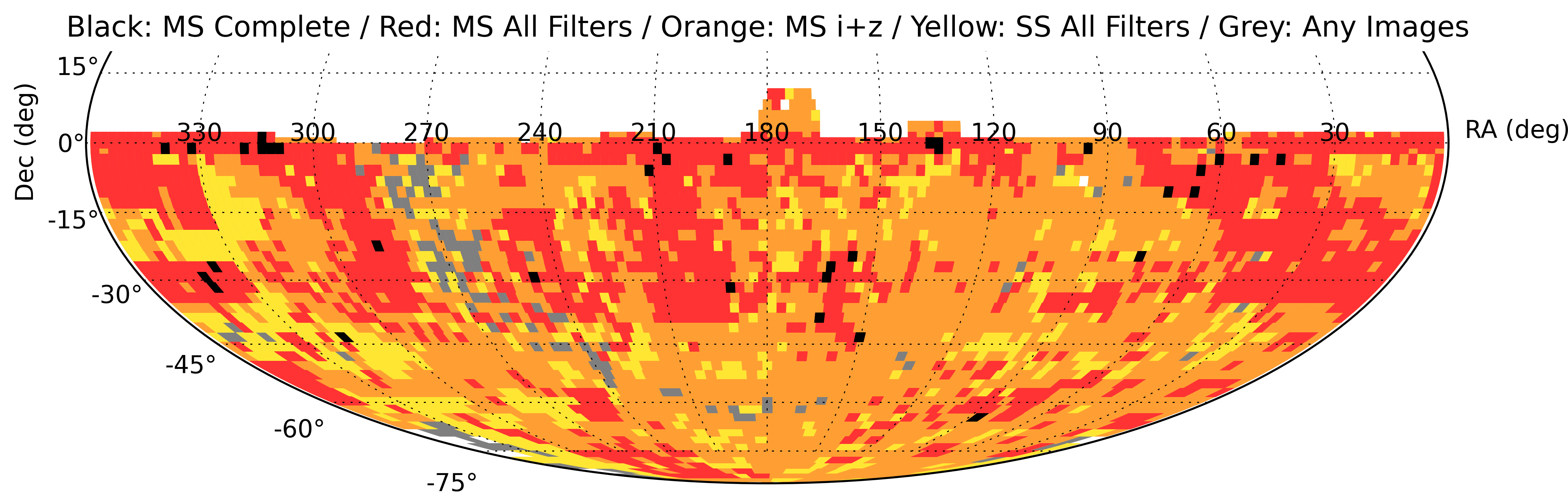}
\caption{Coverage of SkyMapper DR2, colour-coded to indicate the progress on different fields: ({\it black}) complete Main Survey coverage in all six filters; ({\it red}) at least one Main Survey image in all six filters; ({\it orange}) Main Survey images in $iz$ filters; ({\it yellow}) Shallow Survey images in all six filters; ({\it grey}) any images. Main Survey images have exposure times of 100~s in each filter, while the Shallow Survey exposures in $u,v,g,r,i,z$ are 40, 20, 5, 5, 10, 20~s.}\label{fig:coverage}
\end{center}
\end{figure*}

The DR2 Main Survey dataset provides nearly full hemispheric coverage in $i$ and $z$ filters (see Figure~\ref{fig:coverage}), which will serve, among other purposes, as reference frames for the detection of optical transients related to gravitational-wave events detected with Advanced LIGO, such as the kilonova from the binary neutron star merger GW170817 \citep[e.g.,][]{MMOBNS,Andreoni17}. 

This release was made available\footnote{See the SkyMapper website: http://skymapper.anu.edu.au} to Australian astronomers, SkyMapper partners, and their world-wide collaborators on 27 February 2019. The proprietary period is currently expected to last 18 months, after which DR2 will become world-public without restrictions.

This paper is organised as follows: in Section~\ref{sec:enhance}, we describe the Main Survey and changes in processing relative to DR1; in Section~\ref{sec:properties}, we describe properties of the DR2 dataset; in Section~\ref{sec:access}, we provide an update on the data access methods; and in Section~\ref{sec:future}, we discuss the future survey development.

\begin{figure*}
\begin{center}
\includegraphics[width=0.49\textwidth,frame]{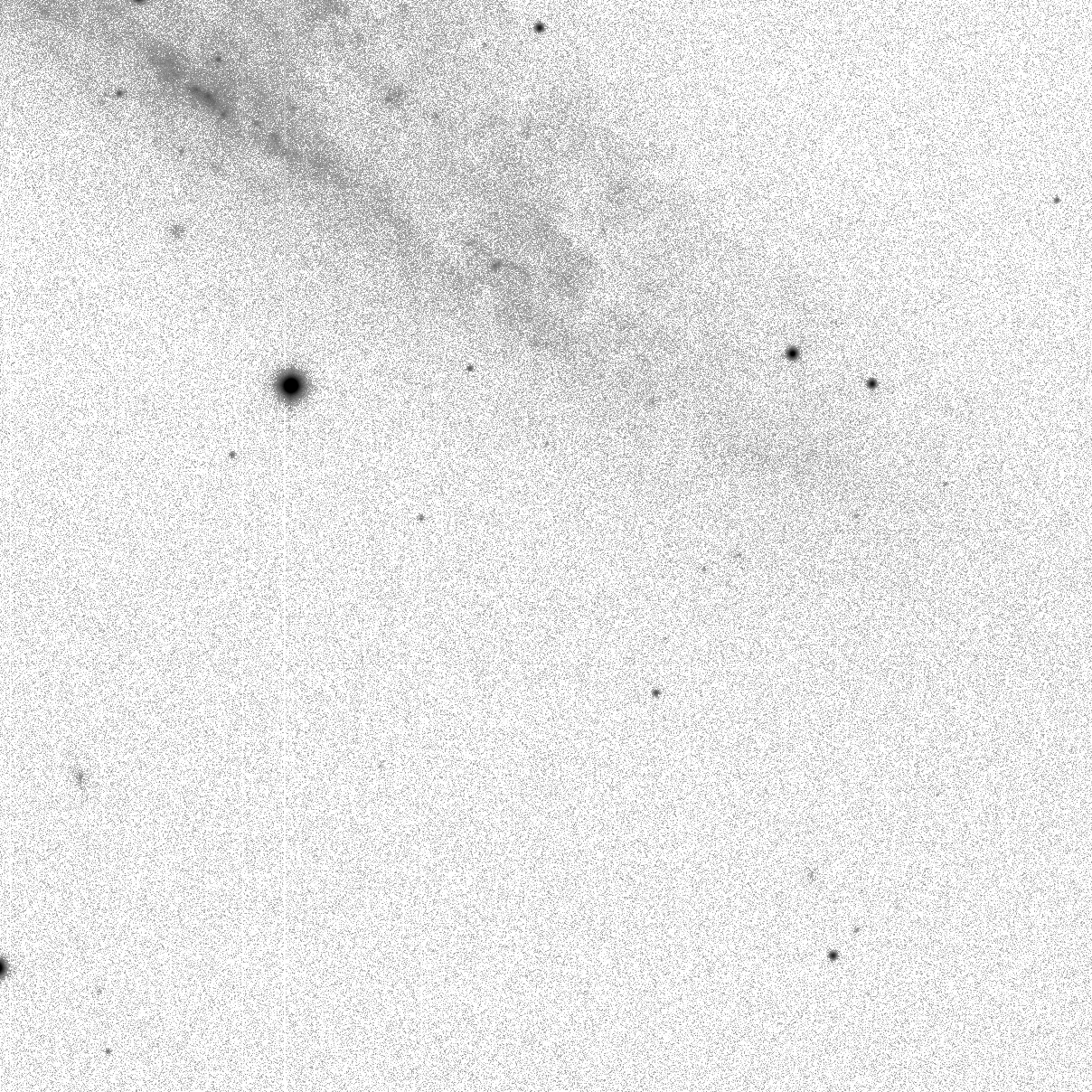}
\includegraphics[width=0.49\textwidth,frame]{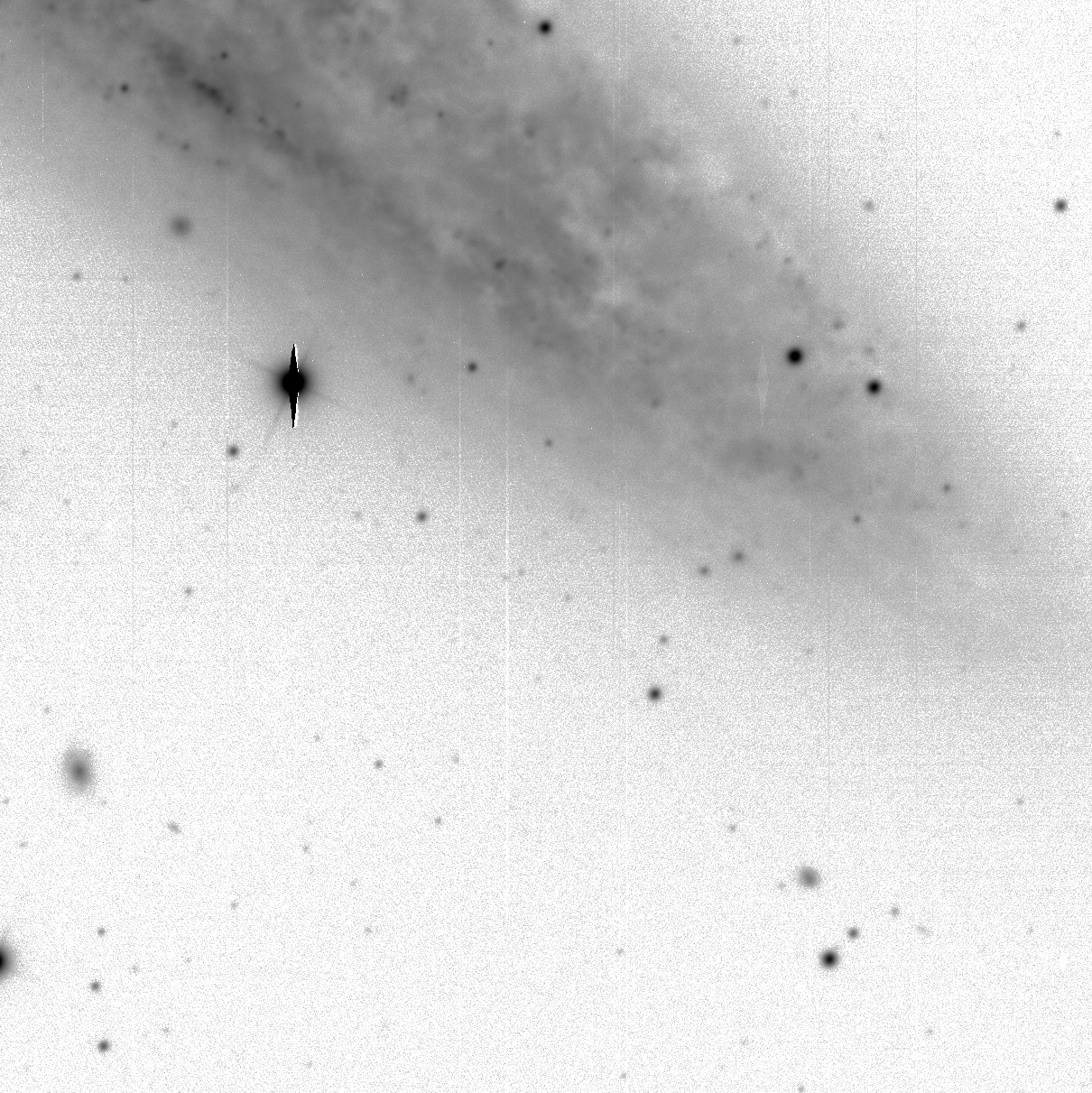}
\includegraphics[width=0.49\textwidth,frame]{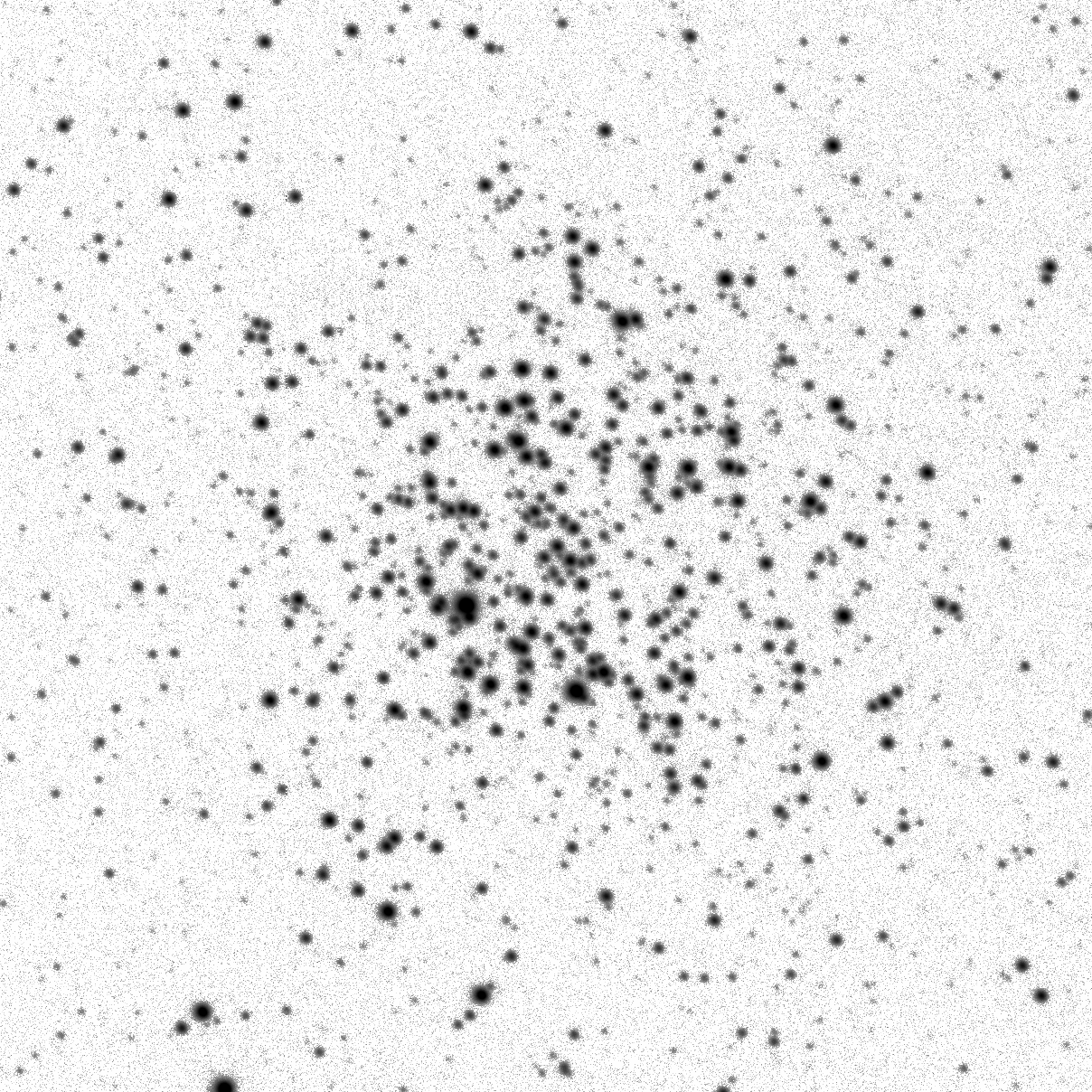}
\includegraphics[width=0.49\textwidth,frame]{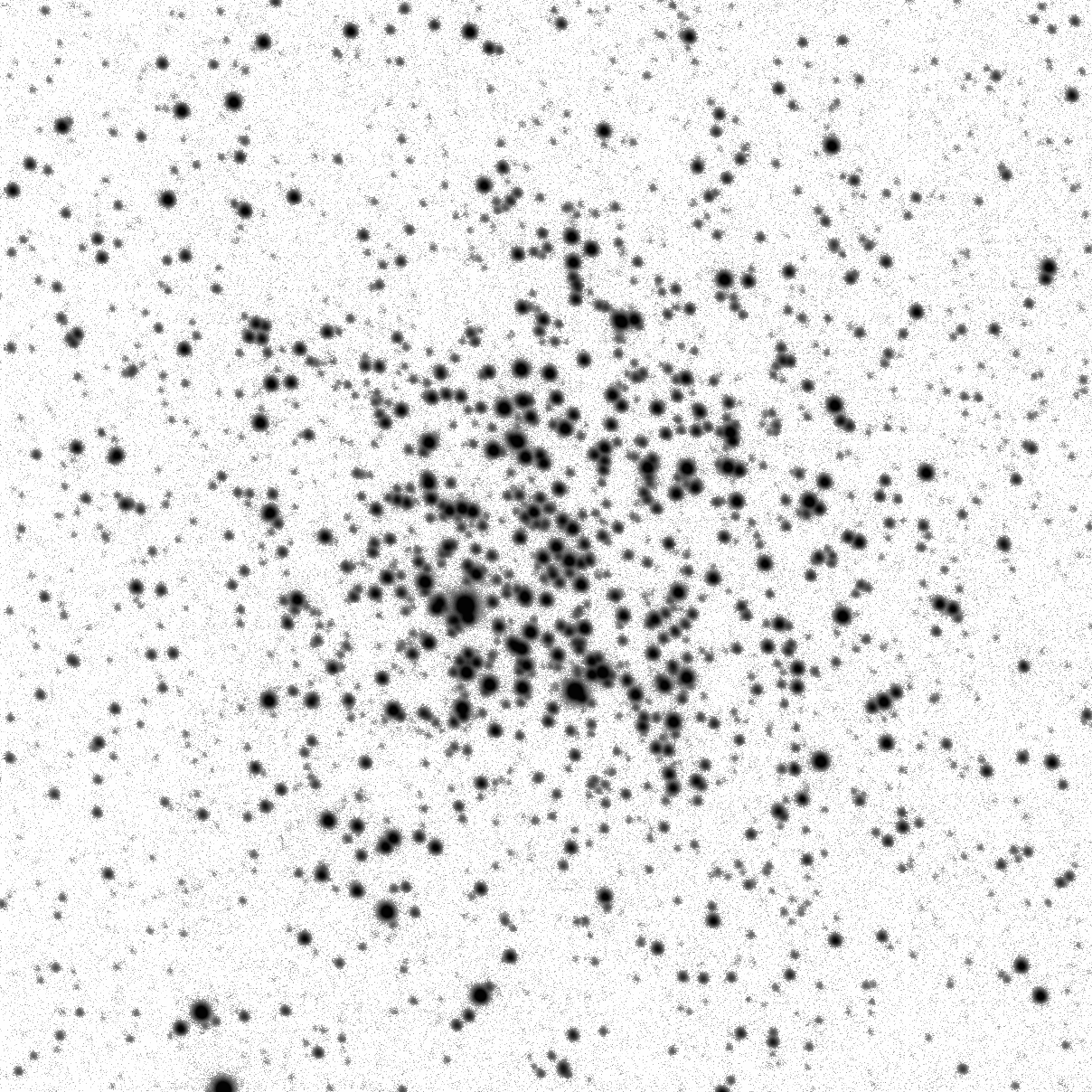}
\caption{Comparison of image pairs from the two survey components (size $10\arcmin \times 10\arcmin$): {\it Shallow Survey} (left, $t_{\rm exp}=5 \ldots 40$~s) and {\it Main Survey} (right, $t_{\rm exp}=100$~s). {\it Top:} $g$-band images of the galaxy NGC~253. {\it Bottom:} $u$-band images of the open cluster Messier~11. While the exposure time ratio in $g$-band is 20, it is only 2.5 in $u$-band.  North is up and East is left.}\label{NGC253}
\end{center}
\end{figure*}

\section{Enhancements in DR2}\label{sec:enhance}

DR2 builds on the Shallow Survey data and processing of DR1 described by \citet{Wolf18a}, and we refer the reader to Section 4 of that paper for the details of the image processing and photometric measurements. The primary gain planned for DR2 was the inclusion of deeper images from the Main Survey, as well as more Shallow Survey data. However, we also improve the data products further (as described in the subsections below) by applying stricter cuts on image quality, by removing fringes in $iz$ images, and by using an improved photometric zeropoint calibration tied to {\it Gaia} DR2.

Main Survey images are always exposed for 100~sec, while the Shallow Survey exposure times range from 5~sec in $g$ and $r$ filters to 40~sec in $u$-band. The gain in depth is often above the naively expected $\sqrt{t_{\rm exp}}$ factor, as the Shallow Survey images are affected by read noise. Figure~\ref{NGC253} compares typical exposures from each survey component side by side, targeting part of the Sculptor galaxy NGC~253 and the open cluster Messier 11. 

While the saturation magnitude for point sources is typically 10~mag in $g$ and $r$ filters in the Shallow Survey, the factor 20 increase in exposure time in the Main Survey moves it fainter by 3.25~mag. However, the completeness limit of point-source detection, which is on the order of 18~mag in the Shallow Survey, is expected to push fainter by less than that due to increased sky background. 

The image dataset for DR2 is just like that of DR1 -- reduced images from each charge-coupled device (CCD) in the mosaic are accessible via an image cutout service on the SkyMapper website, along with the corresponding mask image. Non-zero values in the mask reflect, for each pixel, one or more of the following issues: 1 - non-linear; 4 - affected by saturation; 8 - affected by cross-talk from the other half of that CCD; 32 - a cosmic ray was removed from the pixel; and 64 - affected by electronic noise (see Section~\ref{sec:compnoise}). Masked pixels are treated in two ways during the production of photometry: non-linear pixels are interpolated over, while other issues are captured by {\sc Source Extractor} \cite[version 2.19.5;][]{BA96} into the parameters IMAFLAGS (the OR-combined values of the mask within the isophotal aperture) and NIMAFLAGS (the number of affected pixels within the isophotal aperture).

The structure of the DR2 data tables is very similar to that of DR1, with a few important modifications described in Section~\ref{sec:access}. The list of distinct astrophysical objects, and their averaged primary properties, is contained in the \texttt{master} table. The individual photometric measurements from each image are given in the \texttt{photometry} table. The tables describing the observations themselves are organised into an \texttt{images} table containing image-level properties, a \texttt{ccds} table that incorporates CCD-specific information, and a \texttt{mosaic} table that describes the relationship between the CCDs in the mosaic. A set of external catalogues with DR2 cross-matches is also available. The construction of the data tables is described in more detail in Sections \ref{sec:distill}, \ref{sec:photom_table}, and \ref{sec:im_table}.

\subsection{Observing strategy and cadence}\label{sec:strategy}

The observing strategy in the Shallow Survey is very simple: any time a field is targeted, it is observed with a full colour sequence of six filters that is completed in under four minutes, if there are no interruptions. The Main Survey, in contrast, has several components that are completed over different visits to a field: 
\begin{enumerate}
    \item For the first long visit, we take a colour sequence that includes three exposures for the filters $u$ and $v$ and one exposure for the remaining filters $griz$, using the order $uvgruvizuv$. The sequence is executed as a block and typically completed in 20~min.
    \item Additional exposure pairs in the filters $g$ and $r$ are added during dark and grey time; since March 2017 we restricted the $gr$ pairs to dark time, because $g$-band is our most sensitive filter to moonlight\footnote{The greater bandpass width and superior sensitivity of $g$ compared to $u$ and $v$ overcomes the difference in central wavelength.}. Two $gr$ blocks are collected on different nights and typically executed in 4~min each.
    \item Additional $iz$ pairs are executed as 4-min blocks and observed during astronomical twilight (between $-12\arcdeg$ and $-18\arcdeg$ Sun altitudes) or in bright time, regardless of other progress on the field. Up to three $iz$ pairs are added over the course of the survey.
    \item The final step is a second colour sequence that is identical to the first.
\end{enumerate}

Owing to this strategy, we obtain near-instantaneous six-filter SEDs via the colour sequences, and we gain more depth by adding further data when the observing conditions are most suitable. As a result of this strategy, DR2 contains a mix of completion statistics across the hemisphere. Since 2014 November 14, the astronomical twilight time at SkyMapper has been used exclusively for Main Survey $iz$ pairs, which explains why these filters already provide nearly hemispheric coverage with at least one visit. Nearly 33\% of the Southern sky is covered with all six filters in the Main Survey, typically including one colour sequence, two $gr$ pairs and some $iz$ pairs, i.e., three exposures per filter. The full data set brings the exposure numbers up to $(6,6,4,4,5,5)$ for the $(u,v,g,r,i,z)$ filters but is only complete for 1\% of the hemisphere in DR2.


The resulting cadence of repeat observations also depends on the filter: the colour sequences provide a short-cadence lightcurve in $uv$ consisting of three pairs separated by typically 8~min. Repeat observations of $gr$ pairs and $iz$ pairs take place on separate nights, hence they could be as close as $\sim 20$~hours, or separated by years. The final colour sequence is typically at least a year after the first colour sequence, so the full spectral energy distribution of faint sources is probed only for long-term variability.

Main Survey exposures have a median separation of 14 days in the filters $g$ and $r$, and 1~year in the filters $i$ and $z$. The distribution of time difference between the first and the second colour sequence has two peaks, at 1~year and at 1~month (owing to the Lunar period). 

While the cadence described above holds for a given SkyMapper field, it does not necessarily apply to every object within the field. Because of the gaps between CCDs within the mosaic (0$\farcm 5$ and $3\farcm 2$ between rows, $0\farcm 8$ between columns), a dither pattern is been applied for the observations of each visit. For the Main Survey observations in each filter, the nominal pattern of (RA, Dec) offsets from the field centre is ($-5\arcmin$, $-1\farcm 7$), ($-1\farcm 7$, $+5\arcmin$), ($+1\farcm 7$, $-5\arcmin$), ($+5\arcmin$, $+1\farcm 7$), ($-8\farcm 6$, $-8\farcm 6$), ($+8\farcm 6$, $+8\farcm 6$). The malfunction of various detector controller electronics since October 2017 has led to modifications of the pattern, with step sizes as large as $12\arcmin$, but only 1\% of DR2 data entails such a large offset. For the Shallow Survey, the dither pattern of the first three visits ranges up to $5\arcmin$ from the field centre, but the subsequent visits apply roughly half-field ($1\fdg 1$) offsets in each of RA and Dec, in order to yield more homogeneous photometry from the complete dataset.


\subsection{Image selection}

We started from $\sim$160\,000 images that were observed between 2014 March 15 and 2018 March 14 and processed with our Science Data Pipeline \citep[SDP;][]{ADASS_Luvaul,ADASS_Wolf}. We then selected 121\,494 images using the following quality control criteria:
\begin{enumerate}
\item While the median number of calibrator stars for zeropoint determination is over 1\,600 per frame, we require at least 6 calibrator stars; only 0.6\% of frames have less than 100 stars.
\item We fit linear throughput gradients across our wide field-of-view to the calibrator stars and reject images with strong gradients. Our mean ensemble gradients are consistent with zero; we require that their slopes do not exceed 0.05~mag per $2\fdg 3$ width of the field-of-view in $griz$ filters and no more than 0.1~mag in the $uv$ filters. 
\item We measure the root-mean-square (RMS) scatter of zeropoints among the calibrator stars to identify frames with uneven throughput due to structured cloud or other reasons, and reject frames where the RMS exceeds 0.05~mag in $griz$ filters and 0.12~mag in the $uv$ filters.
\item We determine typical zeropoints per filter, which drift in time as the telescope optics accumulate dust, and reject frames with low throughput, when the loss is on the order of 1~mag or more.
\item We reject frames where the mean point spread function (PSF) has a full-width at half maximum (FWHM) above 5$\arcsec$ or elongation (the ratio of semi-major to semi-minor axis lengths) above 1.4.
\item We tighten the constraints on the World Coordinate System solutions compared to DR1, such that both corner-to-corner lengths of each CCD are required to be within 1$\arcsec$ of the median values for that particular CCD and filter.
\item We reject frames with high background and use thresholds ranging from 500 counts for $uv$ filters, irrespective of exposure time, to thresholds of 3\,000 counts for Main Survey frames in $iz$ filters.
\end{enumerate}

\subsection{Data Processing Differences}

The major improvement in the image processing compared to DR1 is the treatment of fringing in the $iz$ filters. Each filter was investigated with a principal component (PC) analysis to determine the main modes of fringing behaviour \cite[similar procedures have been described by][]{2017JInst..12C5003B,2017PASP..129k4502B}. We expect the fringes to be a linear combination of patterns, each of which is created by groups of night-sky emission lines that vary separately in intensity. The fringe pattern caused by each independently varying group would thus be represented by a PC. We find that this approach of multiple, independent fringe frames suppresses fringe residuals in SkyMapper images better than using only a mean fringe pattern, i.e., the first PC.

Taking about 5\,000 bias- and flatfield-corrected Main Survey images for each filter, the astrophysical objects in each image were removed and a large-spatial-scale background map was created and subtracted. For the final analysis, we chose a subsample of images that were not excessively dense with objects, were free of large numbers of saturated pixels, and had background levels (prior to subtraction) between 250 and 2\,500 counts. About 3\,000 images per filter, selected independently for each CCD, were run through a PC analysis routine \citep[using the {\it scikit-learn} {\sc Python} module;][]{sklearn} to produce the top 20 PCs. For each CCD, the PC creation required $1-2$ hours of processing time and $400-500$~GB of RAM on the \emph{raijin} supercomputer at the National Computational Infrastructure\footnote{http://nci.org.au}. 

\begin{figure}
\begin{center}
\includegraphics[width=\columnwidth]{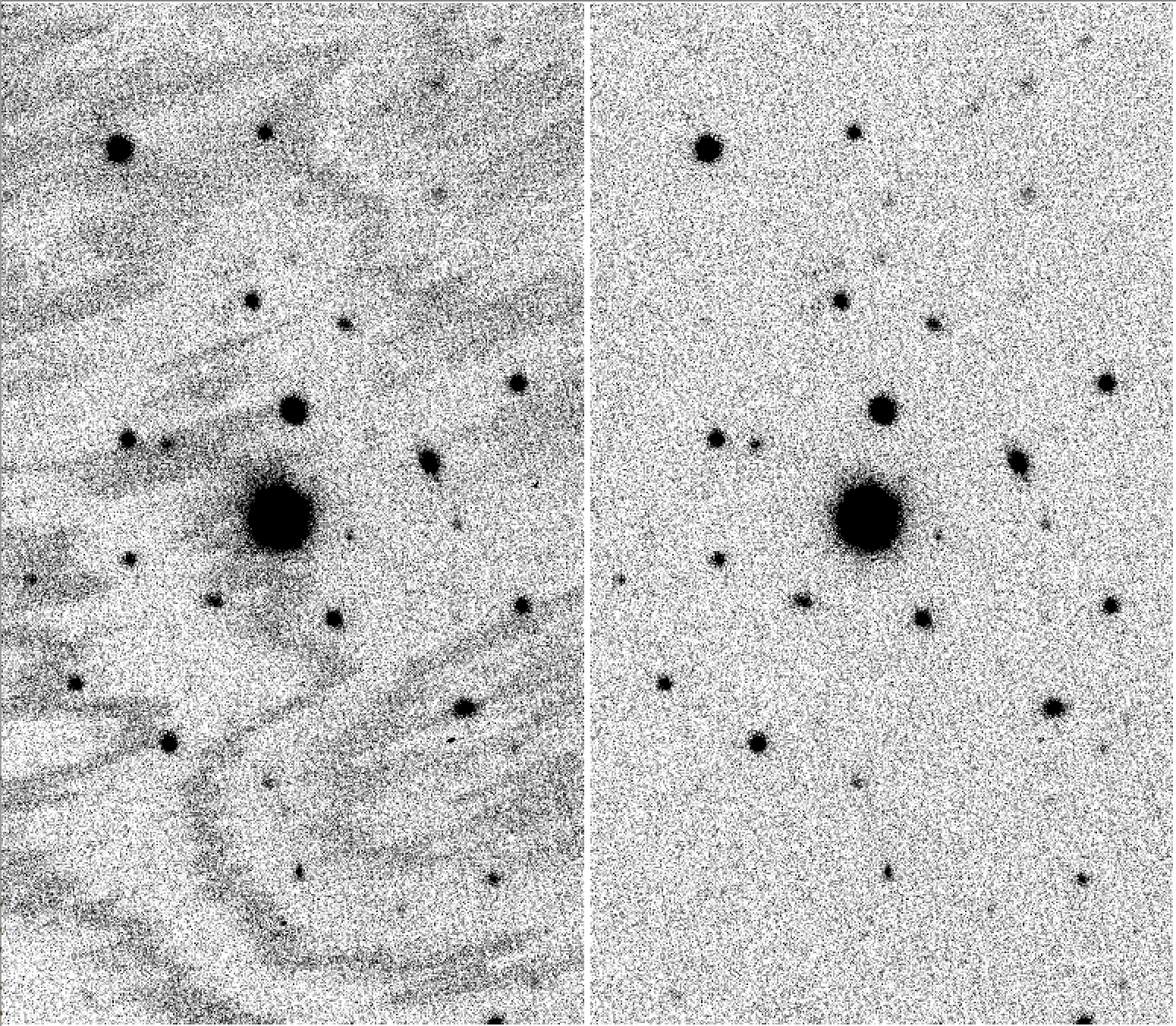}
\caption{Example Main Survey $z$-band image before ({\it left}) and after ({\it right}) defringing with 10 PCs. The peak-to-peak amplitude of the fringes is up to 30~counts in this particular image, with a background level of $\sim$400~counts.}\label{fig:fringe}
\end{center}
\end{figure}

We employed 3 PCs for $i$-band images and 10 PCs for $z$-band images. The choice of the number of PCs to use for each filter was guided by experimentation with fitting different numbers of PCs to a set of test images. This by-eye estimation as to when the fringing pattern was no longer visible against the sky noise turned out to be when the last included PC accounted for about 6\% of the variance explained by the first PC. An example of the method's effectiveness is shown in Figure~\ref{fig:fringe} for a $z$-band Main Survey image. The fitting process mirrored that of the PC generation: astrophysical objects were removed, a large-spatial-scale background map was created and subtracted, and the PCs were fit using the linear algebra least-squares fitter in {\sc NumPy} \cite[{\it linalg.lstsq};][]{numpy}. While the primary goal of fitting the fringes was to clean the Main Survey images, about 4\,200 Shallow Survey images in each of $i$- and $z$-band were also corrected (about 1\,000 new Shallow Survey images in each filter were {\it not} corrected). Any Shallow Survey images that were included in DR1 were not reprocessed (and therefore, not defringed), although new photometric zeropoints were determined (see Section~\ref{sec:zp}). Not defringing Shallow Survey images has a limited effect, as the fringe amplitude is below the read noise of the CCDs. In the future, however, all Shallow Survey images will be re-reduced and defringed.

\subsection{Masking of electronic noise}\label{sec:compnoise}

Following a period of observatory downtime in 2015, a new source of electronic noise became apparent in the raw images. With correlated fluctuations across all amplifiers, amplitudes of tens to hundreds of counts, typically spanning 3 pixels in the $x$ direction, and small positional differences between the four quadrants of the mosaic related to timing offsets during the readout process, the source of this noise was subsequently identified as due to ground loops being introduced to the electronics. The problem was rectified by cabling modifications on 2018 July 25.

\begin{figure}
\begin{center}
\includegraphics[width=\columnwidth,frame]{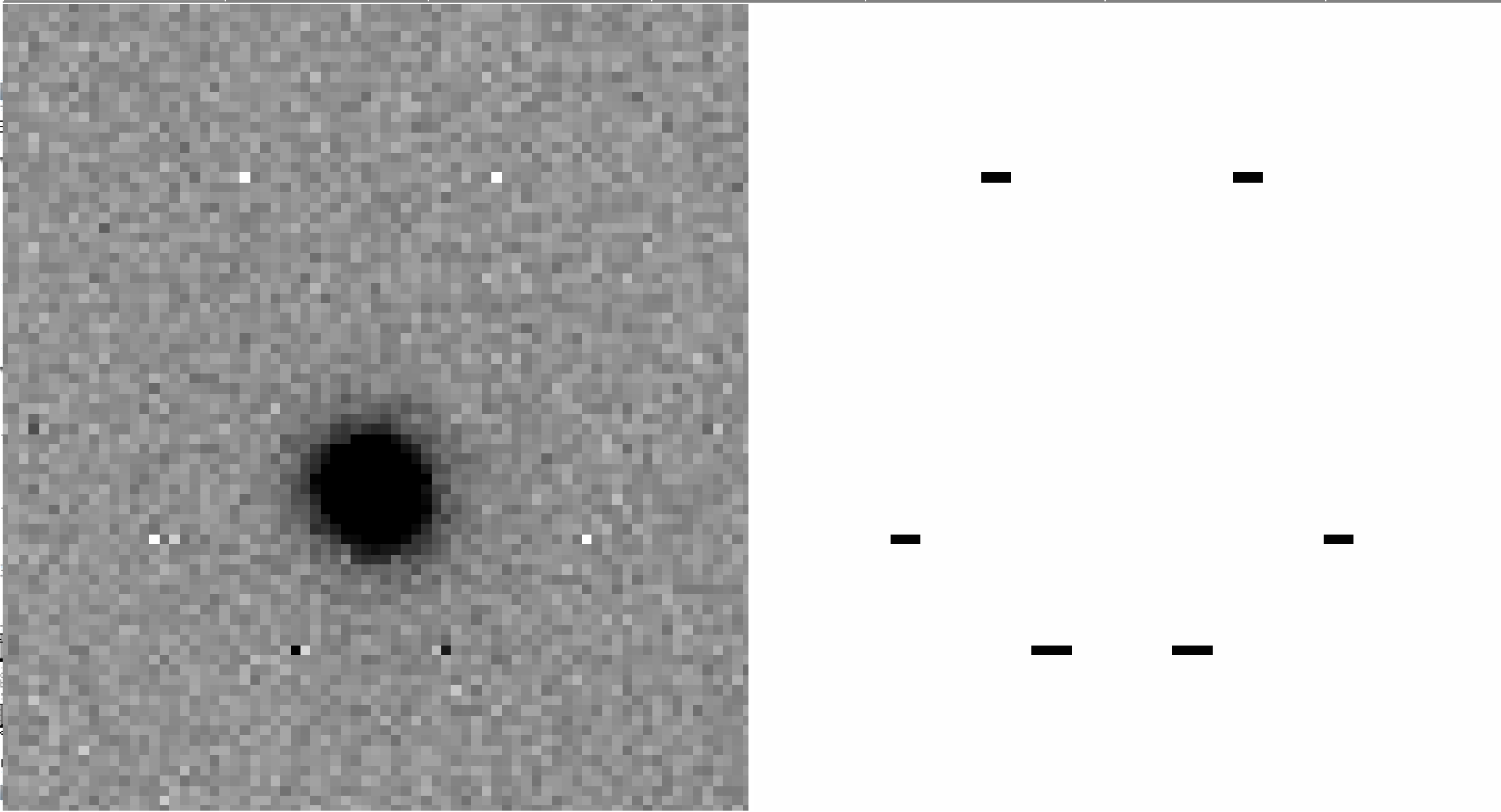}
\caption{Example of electronic noise near the centre of a CCD, showing the near-symmetric behaviour in the two amplifiers used to read out each detector ({\it left}), and the resulting image mask ({\it right}) with the affected pixels flagged by the procedure described in Section~\ref{sec:compnoise}.}\label{fig:compnoise}
\end{center}
\end{figure}

For DR2 images obtained after November 2015, the following procedure was applied to identify and flag the affected pixels in the associated mask file: after flatfield correction and immediately prior to the flagging of cosmic rays, and taking each mosaic quadrant in turn, the astrophysical sources were masked, based on a 2$\sigma$ detection threshold in {\sc Source Extractor}. We also masked the pixels previously flagged by the SDP (bad pixels, saturated pixels, pixels affected by cross-talk, etc.). The data from the 16 amplifiers (eight CCDs) of the quadrant were appropriately flipped to align pixels read out at the same instant, and the minimum unmasked value at each pixel location was used to construct a source-free image. After subtracting the overall median count value of the source-free image to remove the background, pixels with values exceeding 7$\sigma$ in either direction were flagged, as was one pixel to the left and one to the right of the outlying pixel. The flagged pixels were assigned values of 64 in the image mask (see Figure~\ref{fig:compnoise} for an example).

\subsection{Photometric zero-point calibration }\label{sec:zp}

For DR1, we used the American Association of Variable Star Observers Photometric All Sky Survey \cite[APASS DR9;][]{2016yCat.2336....0H} and 2MASS catalogues as the external reference points for photometric calibration, after training a transformation from APASS to Pan-STARRS1 (PS1) bandpasses using stars in a common area and then applying a theoretical colour transformation from PS1 to SkyMapper bandpasses based on stellar templates. We subsequently noted significant inhomogeneities in the APASS zeropoints across the sky, but now that {\it Gaia} DR2 provides all-sky homogeneity, we anchor the SkyMapper DR2 zeropoints to that. 

In the Asteroid Terrestrial-impact Last Alert System (ATLAS) Reference Catalog 2 (Refcat2), \citet{Tonry18} present a transformation from the {\it Gaia} filters $Bp$ and $Rp$ to PS1 $griz$, including a correction based on the flux density of neighbouring stars. This was capable of predicting PS1 $griz$ magnitudes all over the sky to within less than 0.01~mag RMS. We then apply an updated PS1-to-SkyMapper bandpass transformation, using a restricted set of calibrator stars and an updated dust term. The bandpass transformation is derived from synthetic photometry using the stellar spectral library of \citet{1998yCat..61100863P}. 

In the case of the $uv$ filters this process involves extrapolation from the PS1 $g$-band, which has a strong colour term (see below) and is thus prone to significant error propagation. Since we do not have prior knowledge of the metallicity of our zeropoint stars, we ignore the significant metallicity dependence in this transformation, which increases zeropoint scatter due to metallicity scatter among the calibrator stars, and potentially introduces a subtle calibration drift due to stellar-population gradients across the sky. 

Stars are only used as calibrator stars when (i) {\it Gaia} DR2 reports a parallax of $>1$~milliarcsecond, meaning the stars are closer than 1~kpc, to limit the potential impact of dust extinction, (ii) {\it Gaia} photometry is considered reliable and not a blend of sources, and (iii) the star is in a suitable colour range for linear colour terms; for details, see \citet{Tonry18}. We further require that either the integrated \mbox{$E(B-V)$} reddening in the all-sky map of \citet[hereafter SFD]{SFD98} is below 0.3~mag or the star has a meaningful (i.e., non-zero) value for the $A_G$ extinction estimated in {\it Gaia} DR2. 

We estimate extinction levels for sight lines with \mbox{$E(B-V)_{\rm SFD}<0.3$} using \mbox{$A_V = 0.86 \times 3.1 E(B-V)$} \citep{SF11}, while reducing the estimated $A_V$ further at higher SFD dust levels to take into account that the stars are most likely not reddened by the full dust column. To this end, we use the {\it Gaia} reddening estimate $A_G$, but we do not simply adopt it owing to its large noise. Instead, we determine a weighted average between the SFD estimate and the {\it Gaia} estimate, whereby the fractional SFD weight declines from 1 at \mbox{$E(B-V)_{\rm SFD}=0.3$} towards zero for extremely high \mbox{$E(B-V)_{\rm SFD}$} values. Our final bandpass transformations are then (subscript to $u$-band denotes airmass, over which we interpolate when applying to a given image):
\begin{align*}
  u_2 & = g_{\rm PS} + 0.783 + 1.127(g-i) - 0.313\times A_V \\
  u_1 & = g_{\rm PS} + 0.778 + 1.350(g-i) - 0.354\times A_V \\
  v   & = g_{\rm PS} + 0.333 + 1.495(g-i) - 0.459\times A_V \\
  g   & = g_{\rm PS} - 0.012 - 0.174(g-i) + 0.022\times A_V \\
  r   & = r_{\rm PS} + 0.000 + 0.022(g-i) - 0.004\times A_V \\
  i   & = i_{\rm PS} + 0.011 - 0.043(g-i) - 0.007\times A_V \\
  z   & = z_{\rm PS} + 0.016 - 0.043(g-i) - 0.019\times A_V ~. 
\end{align*}

For each SkyMapper image, we match as many stars to the zeropoint catalogue as possible, and fit a two-dimensional plane to the zeropoint values across the mosaic. The resulting zeropoint plane is then applied to all photometry across the image.

Currently, the flatfield process includes no correction for scattered light in the twilight flats, which will be introduced in the future. Heavily dithered observations of the SkyMapper standard star fields indicate that  corrections are mostly less than 2\% in sensitivity, but in the 1\% area of the mosaic that is closest to the corners the calibration offsets can reach locally up to 5\%. This means that an object measured once in a very corner of the mosaic and another time in the inner bulk area could show apparent variability up to a 5\% level in extreme cases.

\subsection{Distill process for \texttt{master} table}\label{sec:distill}

The final step in the production of the data release is a distill process. It uses the \texttt{photometry} table that has one row per detection and creates a \texttt{master} table with one row per unique astrophysical object. When objects have multiple detections in the same filter, these may be from both the Shallow Survey and the Main Survey, and they are combined into best-estimate values assuming the object is not variable. Variable sources are of course better characterised by their individual detections in the \texttt{photometry} table, as sampled by the observed cadence. (The timescales for and between visits were described in Section~\ref{sec:strategy}.) 

\subsubsection{Flagging and combining detections}

We retain any FLAGS values assigned by {\sc Source Extractor} (up to bit 7, value 128) in the \texttt{photometry} table, but can also add several possible values from our post-processing (values 512, 1024, 2048, and 4096): 
\begin{enumerate}
\item As in DR1, we set bit value 512 for very faint detections that we consider dubious and a potential source of error. This includes all sources on Shallow Survey images that have any one of four crucial magnitudes fainter than 19 mag or NULL: the PSF magnitude, the Petrosian magnitude \citep{Petro76}, the 15-arcsec aperture magnitude, and the 5-arcsec aperture magnitude (after applying the aperture correction). Sources on the deeper Main Survey images are flagged when any of the PSF, Petrosian, or 5-arcsec aperture magnitude is NULL, or when the error on the 5-arcsec aperture magnitude is greater than 0.3~mag. This treatment makes sure that very faint detections in the Shallow Survey are ignored for the benefit of letting the Main Survey alone define their distilled properties, while faint Main Survey-only detections are included into the \texttt{master} table. 

\item We set bit value 1024 for sources whose light profile appears significantly more concentrated than a PSF, using the selection rule \mbox{MAG\_APC02$-$MAG\_APR15 $<-1$} AND \mbox{CHI2\_PSF $>10$}; this is unchanged from DR1\footnote{See the SkyMapper website for details of column definitions: http://skymapper.anu.edu.au/table-browser/}.

\item We set bit value 2048 for detections that are too close to bright stars and thus have a high chance to be affected by optical reflections. Around such stars, we flag all sources within a specific angular distance, using an algorithm improved over DR1. We now use bright stars from ATLAS Refcat2, transform their PS1 photometry to SkyMapper $uvgriz$ magnitudes, and calculate the flagging radius around each star as $10^{-0.2 m}$ degrees using the expected SkyMapper magnitude. For stars fainter than $(u,v,g,r,i,z)=(4,5,8.5,8.5,6,5)$ we do not seem to get bad detections and thus do not apply any flags near these. Note that reflections are not concentric around the stars, so the flagging radius has been enlarged to account for the maximum affected area across the whole field-of-view.

\item We set bit value 4096 for a small set of \texttt{master} table entries around the RA=0/360 boundary that were inadequately merged between filters.

\end{enumerate}

When a detection of a source is identified by {\sc Source Extractor} as saturated or affected by many masked pixels, or has been flagged with one of the rules above (criterion: \mbox{FLAGS$>=$4 OR NIMAFLAGS$>=$5}), we declare it as a bad detection, and ignore it when determining \texttt{master} table properties for the source. When there are only bad detections for a source, then, to avoid omitting the source from the \texttt{master} table, we use the bad detections to populate just the position-related columns (including cross-matching with other catalogues), and the photometric columns are set to NULL. Objects without good detections in a given filter can be identified by having \mbox{\{F\}\_NGOOD$=$0}, where \{F\} is the filter name; and if no filters have good detections, then \mbox{NGOOD$=$0}.

We follow the same approach as in DR1 to merge individual detections into parent objects, first within a filter, then among the filters. Sometimes this process creates parent objects with multiple child objects and we treat them as described below. We acknowledge that this process is intrinsically problematic, and hope to improve it for the next data release.

When merging individual detections for one filter, some point sources on some images are split into two child objects due to suboptimal settings in {\sc Source Extractor}. In DR1 we added their fluxes in the distill process, but as we have more observations per source in DR2, we chose to remove such detections from the \texttt{photometry} table and use only the others. We note that we lose an object entirely in a filter when it has multiple children in all images of that filter. This also happens to genuine binary stars when they are merged into a single object by the process. Unfortunately, in DR2 this data cannot be recovered from the provided tables.

When merging the six filters, we encounter situations (as in DR1) where a global object ID is associated with two detections in some of the filters. We then leave out the (likely unphysical) results for those filters from the \texttt{master} table. However, data for an omitted filter can still be found in the \texttt{photometry} table.

\subsubsection{Combining measurements into distilled values}

How we combine individual measurements depends on the nature of the measurement:
\begin{enumerate}
\item For RA/Dec positions and the observing epoch, we determine simple averages and standard deviations of all values; the Petrosian radius, selected only from the $r$-band measurements, is also averaged.
\item For FLUX\_MAX and CLASS\_STAR, we pick the highest value; FLAGS are bitwise OR-combined, and NIMAFLAGS values are summed up.
\item PSF and Petrosian magnitudes are combined with a more complex algorithm: first we determine more realistic errors of our individual measurement values by quadratically adding floors of 0.01~mag to reflect flatfield uncertainties (consistent across all filters); then we use the set of PSF magnitudes to identify outlier measurements: we calculate an inverse variance-weighted median magnitude from the list of values; next we obtain their median absolute deviation (MAD) from the weighted median; then we clip possible outlier measurements from the list when they deviate from the weighted median value by $>3\times$ the larger of the MAD and their individual error. In a final step, we combine the remaining values into an inverse variance-weighted measurement of the mean magnitude and its final error, using the same list of detections for both the PSF and the Petrosian magnitude. The \texttt{photometry} table contains a column USE\_IN\_CLIPPED that identifies the measurements used for the mean. As a variability indicator, we also determine a reduced $\chi^2$ from the full set of PSF magnitudes and store it in column \{F\}\_RCHI2VAR, where \{F\} is the filter name.
\end{enumerate}

Objects that are saturated or have bad flags in all available frames will only have moderately useful information. Several columns are filled with NULL values on purpose, but positions are still averaged and FLAGS are still OR-combined. They can be easily identified by their FLAGS value in the \texttt{master} table ($\ge 4$).

\subsubsection{Cross-matched external catalogues}

As in DR1, we cross-match the \texttt{master} table with several external catalogues. For large photometric catalogues, we determine the matching external object and record its ID and projected distance within the \texttt{master} table. For small catalogues (mostly spectroscopic), we match in reverse direction and record the SkyMapper object and distance in the external catalogue as DR2\_ID and DR2\_DIST. The maximum distance for all cross-matches is $15\arcsec$, motivated by the region in which our 1D PSF magnitudes may be affected.

The cross-matched large catalogues include 2MASS Point Source Catalog (PSC) \citep{2MASS}, AllWISE \citep{WISE, 2011ApJ...731...53M}, ATLAS Refcat2 \citep{Tonry18}, {\it Gaia} DR2 \citep{Gaia18}, the Revised Catalog of {\it GALEX} Ultraviolet Sources \cite[GUVcat;][]{2017ApJS..230...24B}, PS1 DR1 \citep{Chambers16, Magnier16}, SkyMapper DR1 \citep{Wolf18a} and UCAC4 \citep{Zacharias13}. In the case of {\it Gaia}, we record the two nearest matches, which helps with identifying blended sources. The \texttt{master} table is also cross-matched against itself to identify the ID and distance to the nearest neighbour of every source (up to $15\arcsec$). 

The reverse-cross-matched catalogues include the 2dF Galaxy Redshift Survey \cite[2dFGRS;][]{2001MNRAS.328.1039C}, the 2dF QSO Redshift Survey \cite[2qz6qz;][]{2004MNRAS.349.1397C}, the 2dFLenS Survey \citep{Blake16}, the 2MASS Redshift Survey \cite[2MRS;][]{2012ApJS..199...26H}, the 6dF Galaxy Survey \cite[6dFGS;][]{2004MNRAS.355..747J, 2009MNRAS.399..683J}, the GALAH survey DR2.1 \citep{2018MNRAS.478.4513B}, MILLIQUAS v6.2b \citep{2015PASA...32...10F}, the Hamburg/ESO Survey for Bright QSOs \cite[HES QSOs;][]{2000A&A...358...77W} and the AAVSO International Variable Star Index \citep[VSX;][]{VSX06,VSXcat}. When using cross-matched IDs, care needs to be taken to observe the distance column in order to only select detections that are likely to be physically associated.

\section{DR2 Properties}\label{sec:properties}

\begin{table*}
\caption{DR2 image numbers by filter and survey segment. The row {\it All} refers to the {\it sum} for the Images, CCDs, and Detections columns, but for the Fields column refers to the {\it distinct number} that have coverage in all filters. PSF represents the median FWHM in arcsec. The saturation limit in the last column is the median value, but ranges over more than $\pm 1$~mag within each filter due to variable observing conditions.}
\label{tab:imcount}
\centering
\begin{tabular}{l|rrr|rrr|r|rr}
\hline\hline
 & \multicolumn{3}{c|}{Shallow Survey} & \multicolumn{3}{c|}{Main Survey} & \multicolumn{3}{c}{Combined}\\
\cline{2-4}\cline{5-7}\cline{8-10}
Filter & Fields & Images & CCDs & Fields & Images & CCDs & Detections & PSF & $m_{\rm sat}$\\
\hline
$u$ & 3,853 & 13,427 & 393,990 & 1,526 & 5,109 & 159,232 & 115,597,851 & $3\farcs1$ & 9.1 \\
$v$ & 3,721 & 13,032 & 381,379 & 1,486 & 5,065 & 159,424 & 125,643,376 & $2\farcs9$ & 8.7 \\
$g$ & 4,024 & 14,544 & 450,336 & 1,610 & 4,101 & 127,654 & 661,342,772 & $2\farcs6$ & 9.7 \\
$r$ & 4,024 & 14,594 & 452,068 & 1,624 & 4,315 & 133,704 & 889,397,411 & $2\farcs4$ & 9.9 \\
$i$ & 4,022 & 14,638 & 454,251 & 3,545 & 8,882 & 275,996 & 1,475,889,366 & $2\farcs3$ & 10.0 \\
$z$ & 4,024 & 14,711 & 455,626 & 3,595 & 9,076 & 283,366 & 1,446,386,015 & $2\farcs3$ & 10.0 \\
\hline
All & 3,615 & 84,946 & 2,587,650 & 1,424 & 36,548 & 1,139,376 & 4,714,256,791 & $2\farcs6$\\
\hline
\end{tabular}
\end{table*}

\begin{figure}
\begin{center}
\includegraphics[angle=270,width=0.8\columnwidth]{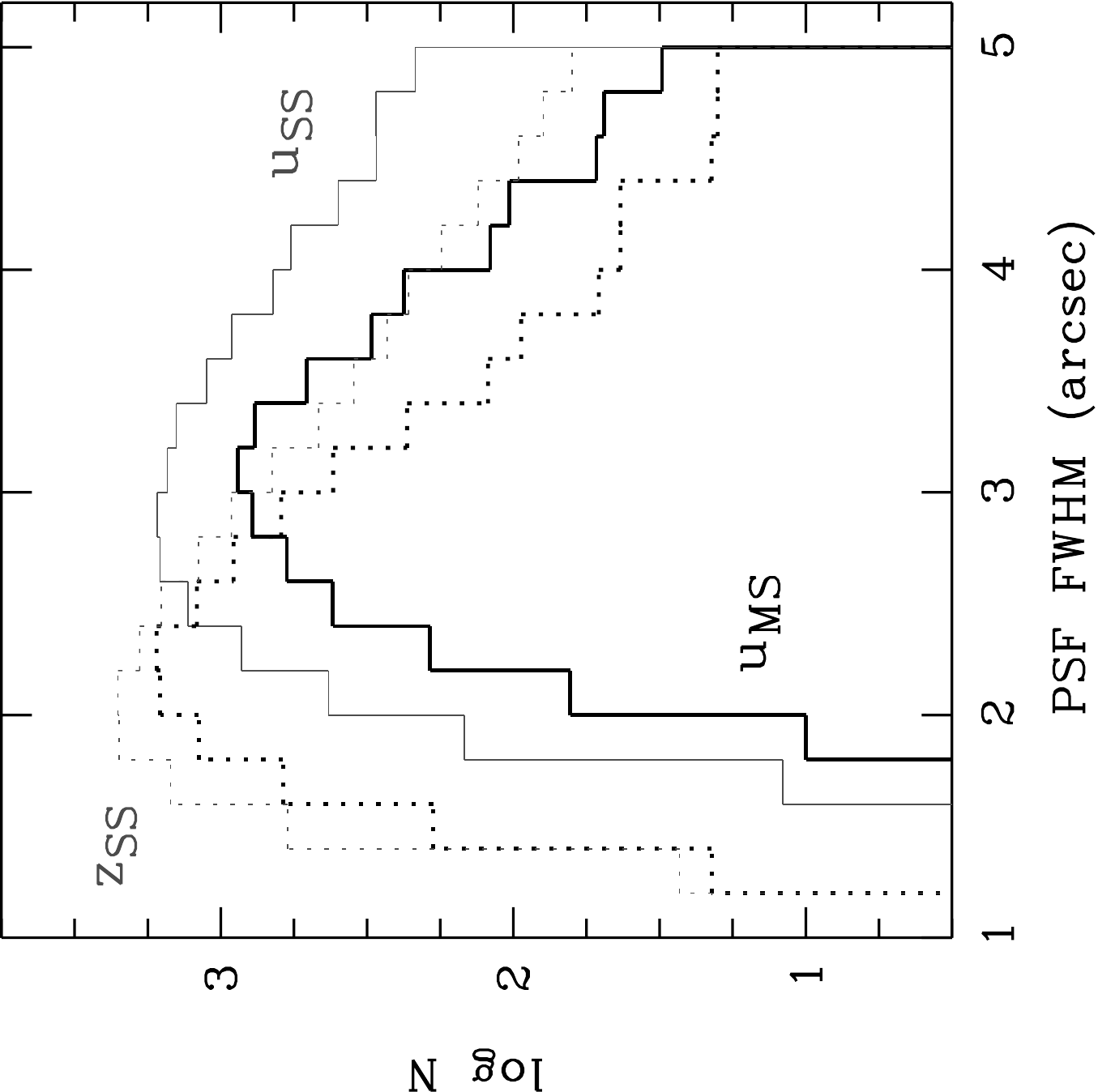}
\caption{Distribution of PSF FWHM in DR2 images: $u$-band images (solid lines) have a median seeing of $3\farcs1$ seeing compared to $2\farcs3$ in $z$-band (dotted lines). The Main Survey (black lines) has a tighter distribution than the Shallow Survey (grey lines).}\label{fig:FWHM}
\end{center}
\end{figure}

\begin{figure}
\begin{center}
\includegraphics[angle=270,width=\columnwidth]{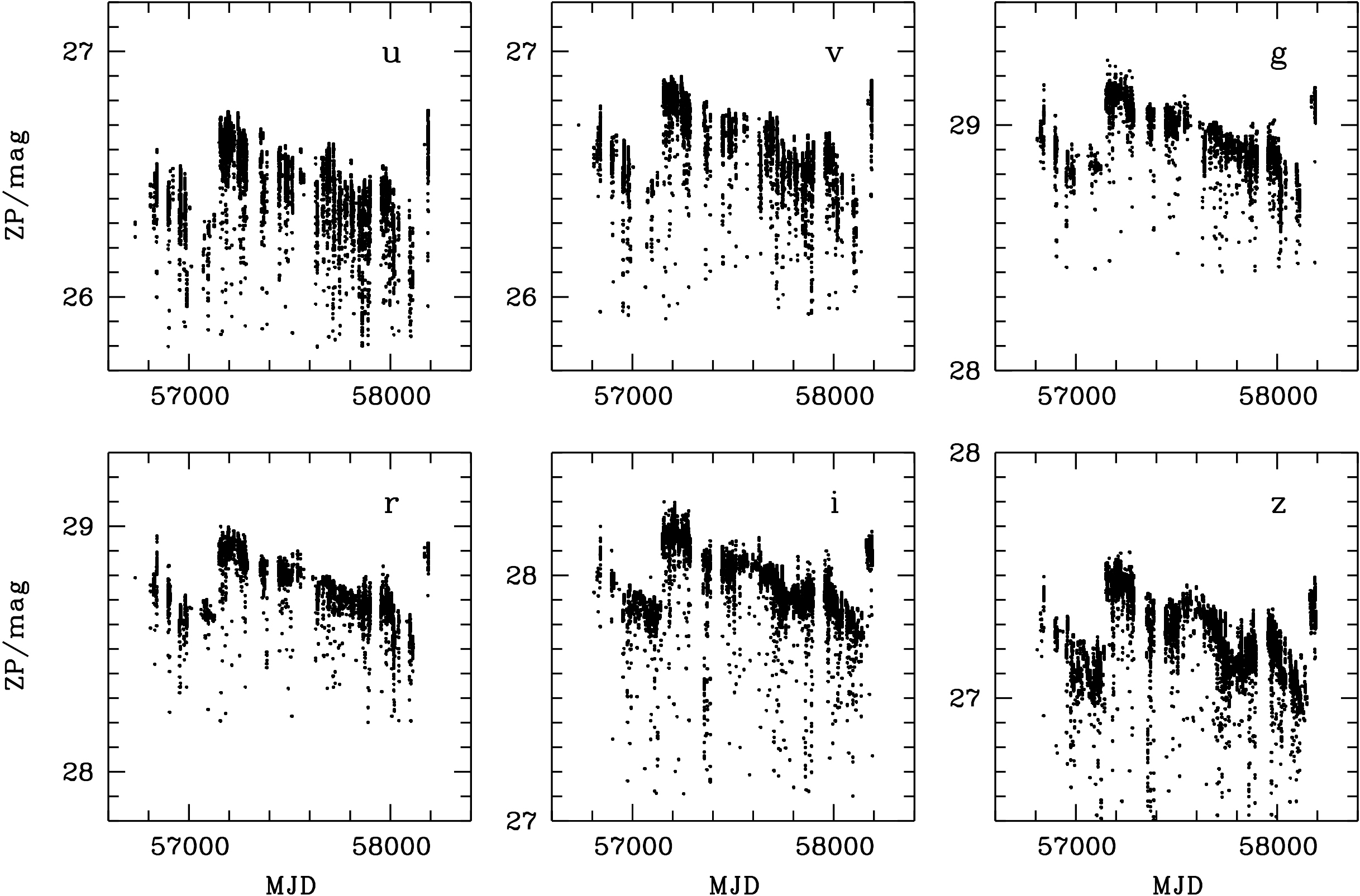}
\caption{Time evolution of magnitude zeropoints per filter for Main Survey exposures: the gradient shows gradual deterioration of optical reflectivity at a mean rate of $-0.33$~mmag/day. Cleaning of the telescope optics on 2015~May~5 (MJD 57147) and 2018~February~1 (MJD 58150) improved the zeropoints by $\sim0.3$~mag each time.}\label{fig:ZP_date}
\end{center}
\end{figure}

\begin{figure}
\begin{center}
\includegraphics[angle=270,width=\columnwidth]{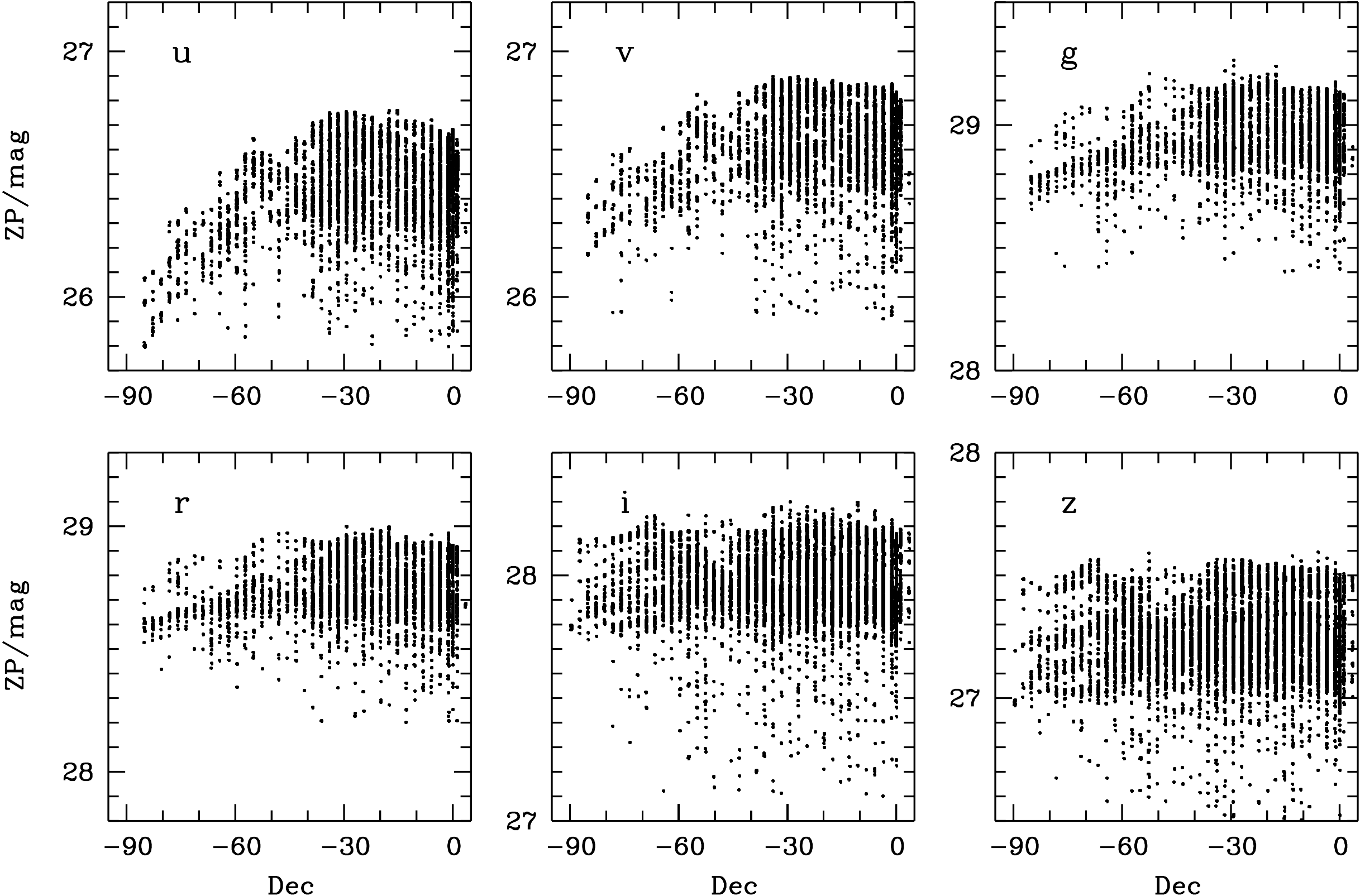}
\caption{Declination dependence of Main Survey zeropoints: the upper envelope results from airmass-dependent atmospheric throughput. The $u$-band suffers $>0.75$~mag loss from zenith to the Celestial pole. While the SkyMapper scheduler attempts to observe at minimal airmass, near-polar fields never rise to low airmass. }\label{fig:ZP_dec}
\end{center}
\end{figure}

\begin{sidewaysfigure*}
\begin{center}
\includegraphics[width=0.49\textwidth]{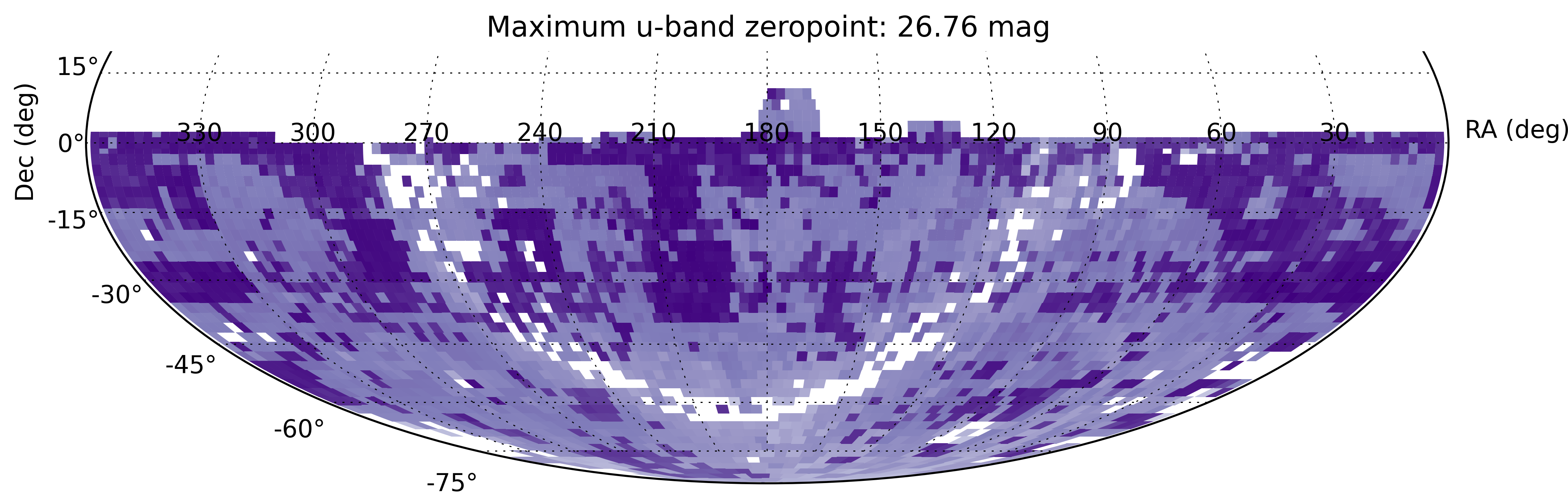}
\includegraphics[width=0.49\textwidth]{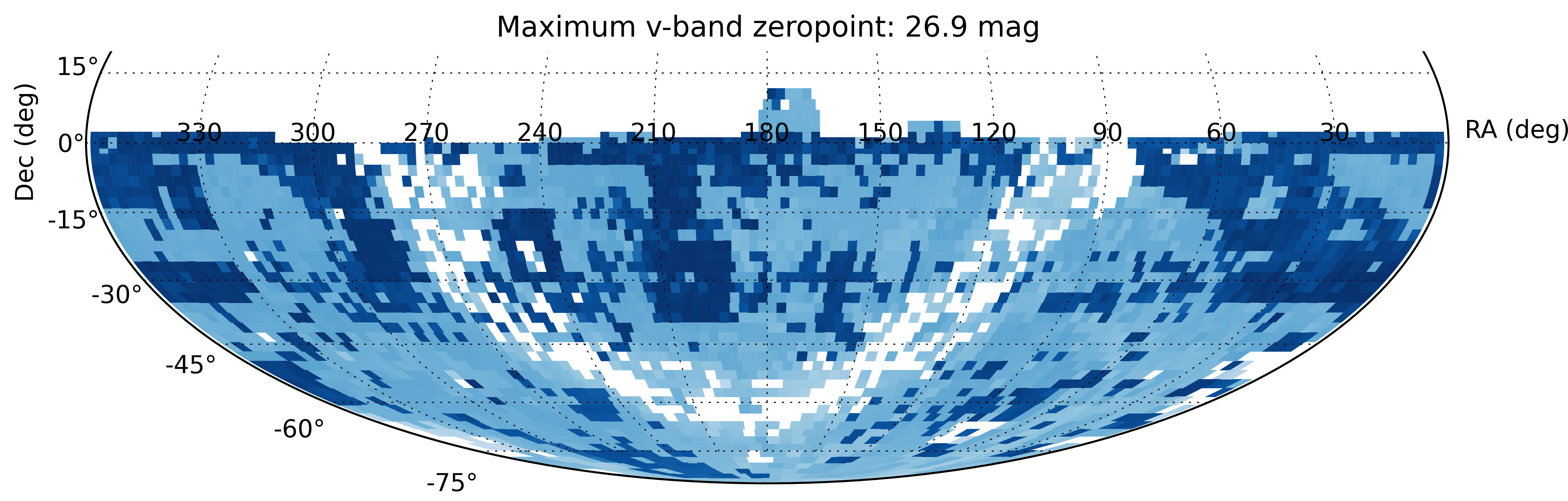}
\includegraphics[width=0.49\textwidth]{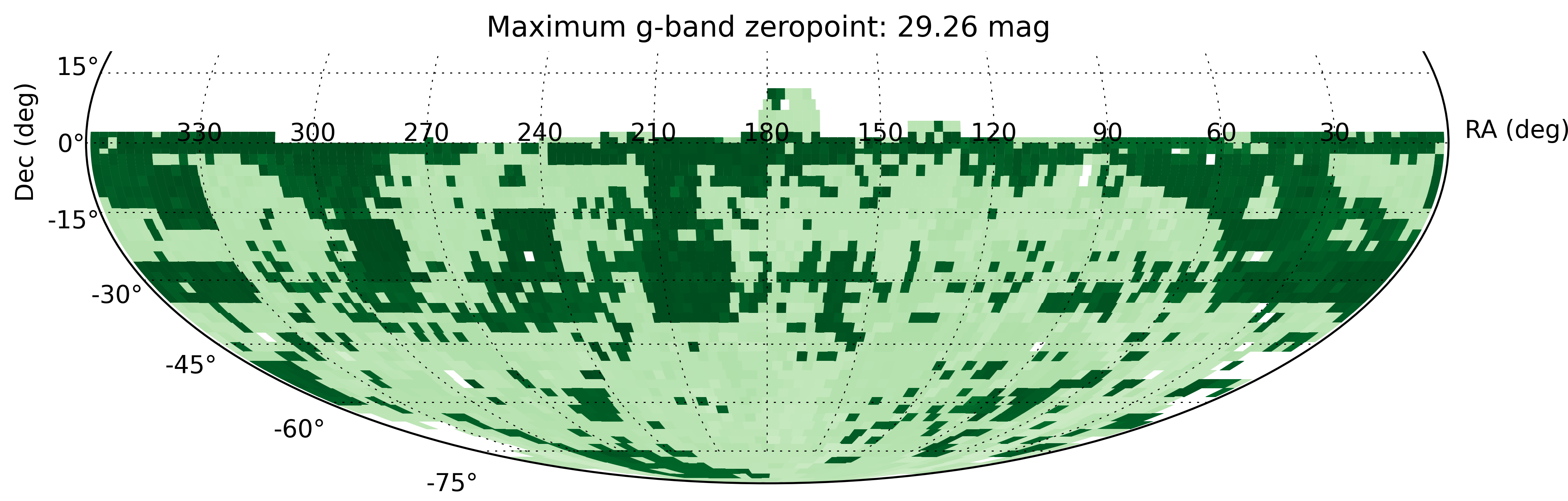}
\includegraphics[width=0.49\textwidth]{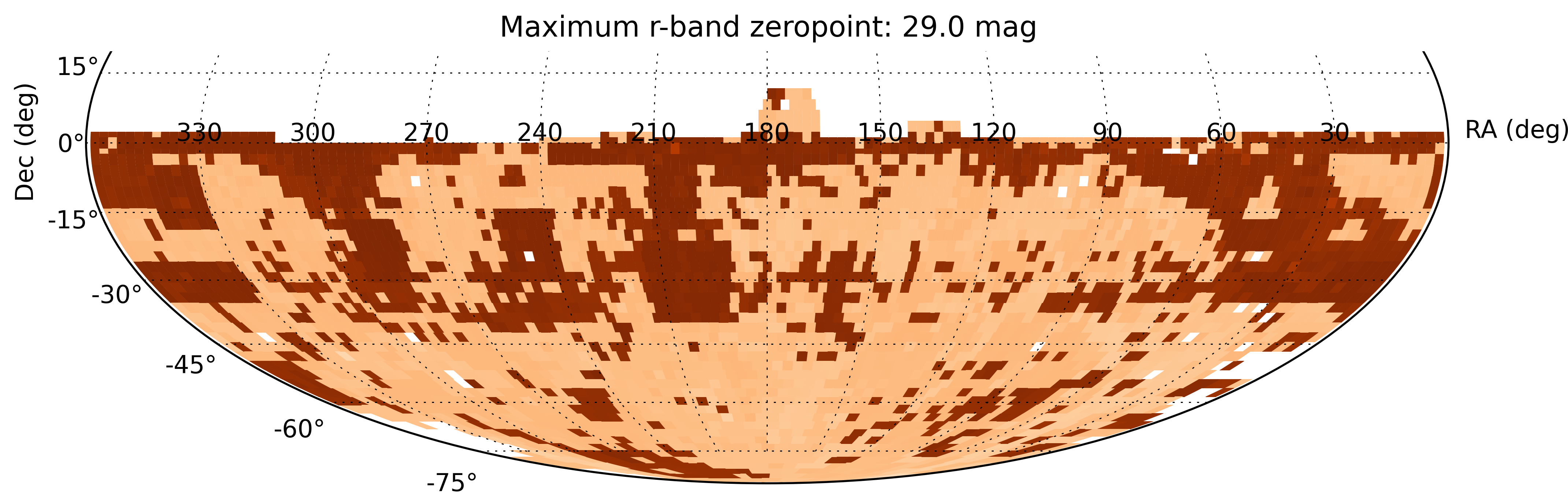}
\includegraphics[width=0.49\textwidth]{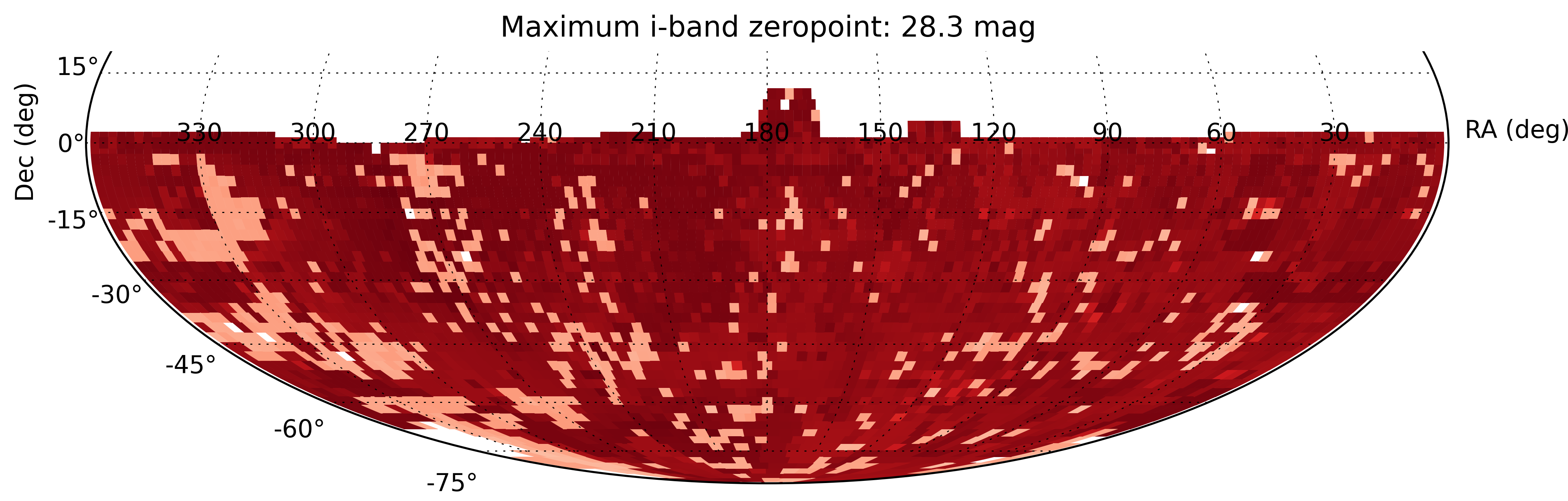}
\includegraphics[width=0.49\textwidth]{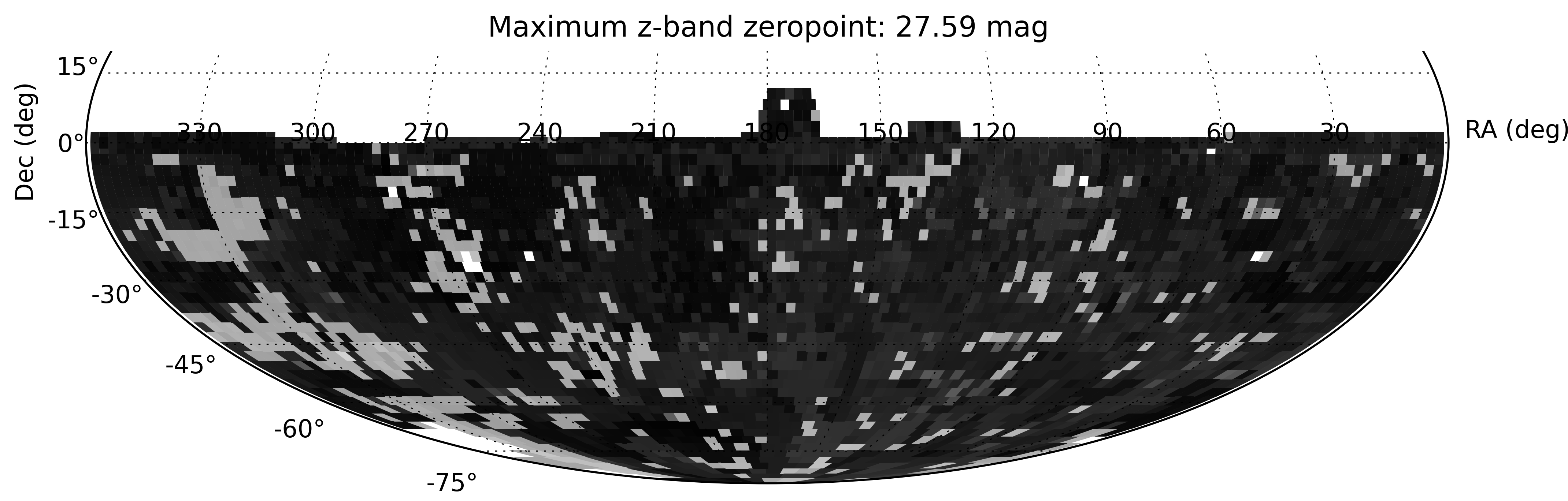}
\caption{Maximum photometric zeropoint for each field, separated by filter. Darker colours indicate deeper data, with white fields having no images. The nominal zeropoint difference between Shallow Survey and Main Survey is roughly given by the ratio of exposure times, amounting to (1.0, 1.75, 3.25, 3.25, 2.5, 1.75)~mag in ($u,v,g,r,i,z$), respectively.}\label{fig:cov6}
\end{center}
\end{sidewaysfigure*}

\subsection{Image dataset}

Data Release 2 contains a total of 121\,494 exposures, comprised of 84\,946 exposures from the Shallow Survey that provide nearly full hemispheric coverage in all filters with multiple visits, and 36\,548 exposures from the Main Survey providing partial coverage in terms of fields, filters, and depth (Table~1). A total of $\sim 3.7$ million individual CCDs passed the quality cuts. The median airmass is 1.11, but a tail to airmass 2 is unavoidable given that the survey footprint includes the South Celestial Pole. 

As in DR1, the median FWHM of the PSF among all DR2 images ranges from $2.3\arcsec$ in $z$-band to $3.1\arcsec$ in $u$-band (see Figure~\ref{fig:FWHM}). Although the median is similar in the two survey components, the distribution is broader in the Shallow Survey: it benefits from occasional short ($<20$~sec) periods of good seeing, but it also includes observations in bad seeing, when the image scheduling software ("the scheduler") avoids the Main Survey and defaults to the Shallow Survey. The median elongation is independent of filter, with values of 1.12 and 1.14 in the Shallow and Main Survey, respectively. 

Most images of the Shallow Survey are read-noise limited, as the median sky background is less than 100 counts, except for the $z$-band images that are mostly background-limited. Median count levels in the Shallow Survey have slightly increased in DR2 as most of the frames added since DR1 were taken in full-moon conditions. In the Main Survey, 98\% of $uv$ frames are read-noise limited because the Main Survey colour sequence is not observed in bright time, and so the median background is 11 counts. The $griz$ frames, in contrast, are all background-limited in the Main Survey, with sky levels always above 100 counts and median levels range from 227 in $g$-band to 646 in $z$-band.

The photometric zeropoints show long-term drifts towards lower efficiencies across all filters due to dust settling on the telescope optics over time (see Figure~\ref{fig:ZP_date}). A cleaning of the telescope optics on 2015 May 5 (Modified Julian Date 57147) improved the zeropoints by $\sim 0.3$ mag, as did another cleaning on 2018 February 1 (MJD 58150). We observe an average trend in the loss of system throughput on the order of 1\% per month, which might motivate an annual cleaning in the future. Most zeropoints for a given filter scatter within 0.1~mag RMS at a given calendar epoch, but some nights with bad weather produce tails with higher atmospheric extinction.

An important influence on zeropoints is airmass-dependent atmospheric extinction. The SkyMapper scheduler attempts to observe any field close to meridian, but for fields with near-polar declination that still means high airmass. This effect translates into a declination-dependent upper envelope for the zeropoints, where fields around $-30^\circ$ enjoy best transparency, while zeropoints deteriorate towards the South Celestial pole (see Figure~\ref{fig:ZP_dec}). The effect is most pronounced in $u$-band (which even changes its filter curve due to the atmospheric cutoff), where we lose over 0.7~mag between zenithal declinations and the pole.


\begin{figure*}
\begin{center}
\hfill
\includegraphics[width=0.49\textwidth]{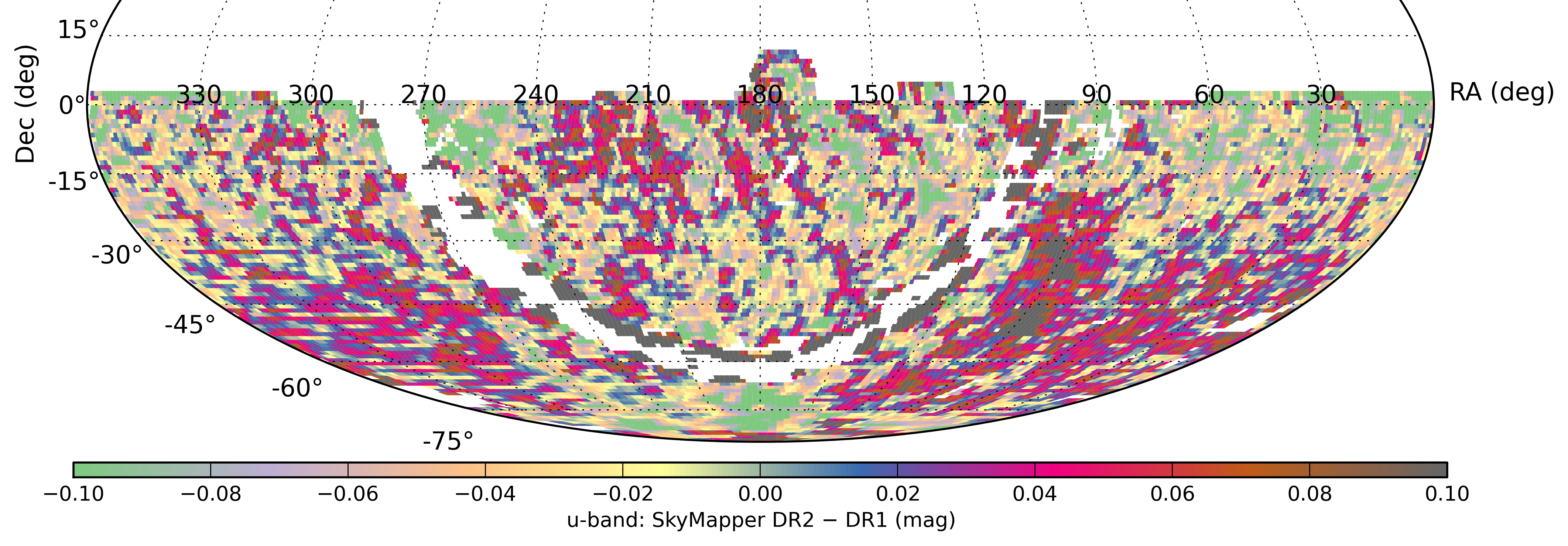}
\includegraphics[width=0.49\textwidth]{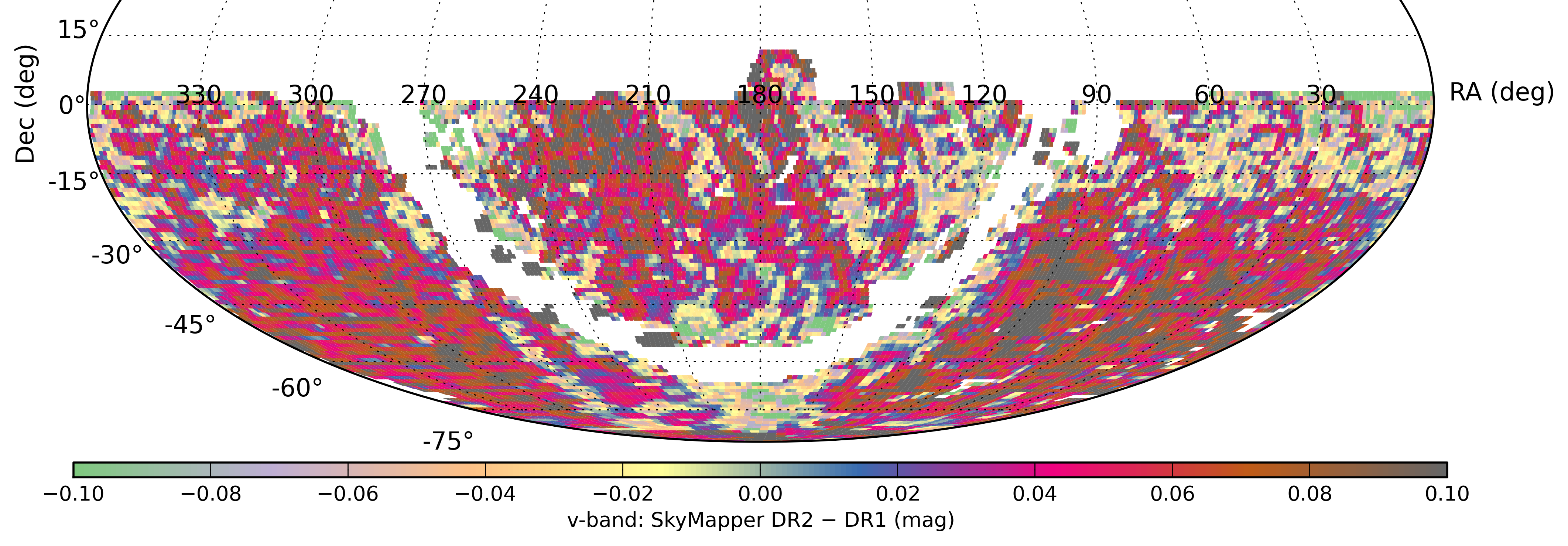}
\includegraphics[width=0.49\textwidth]{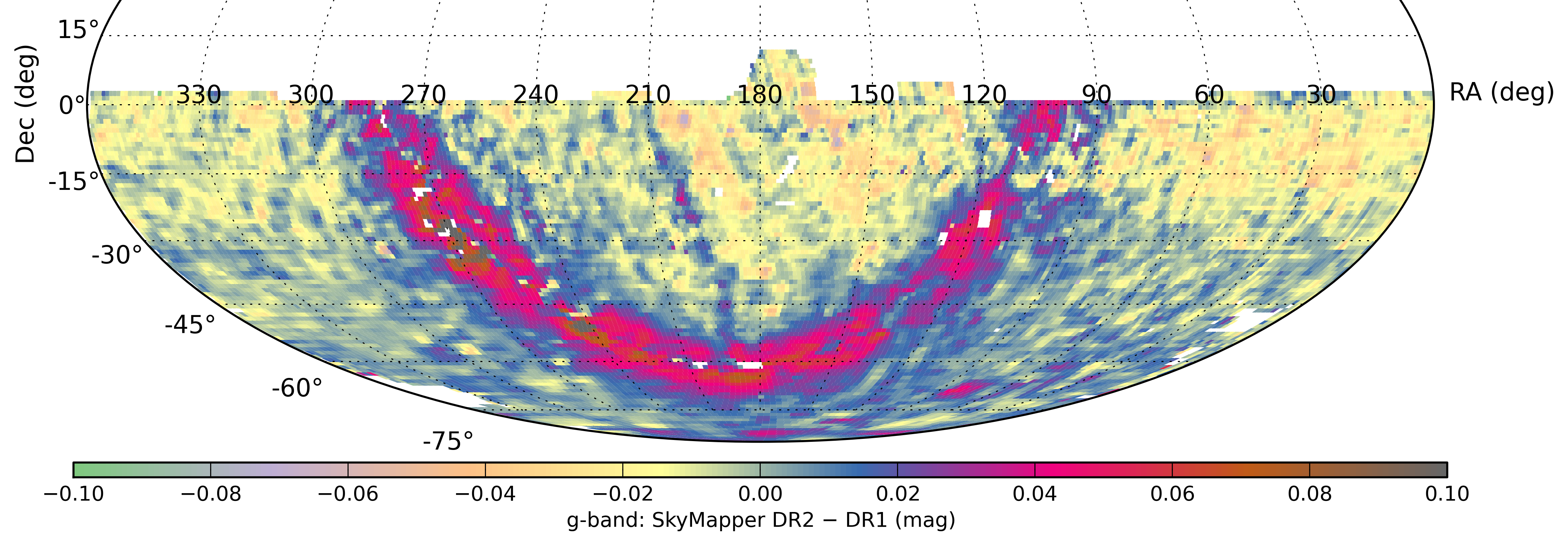}
\includegraphics[width=0.49\textwidth]{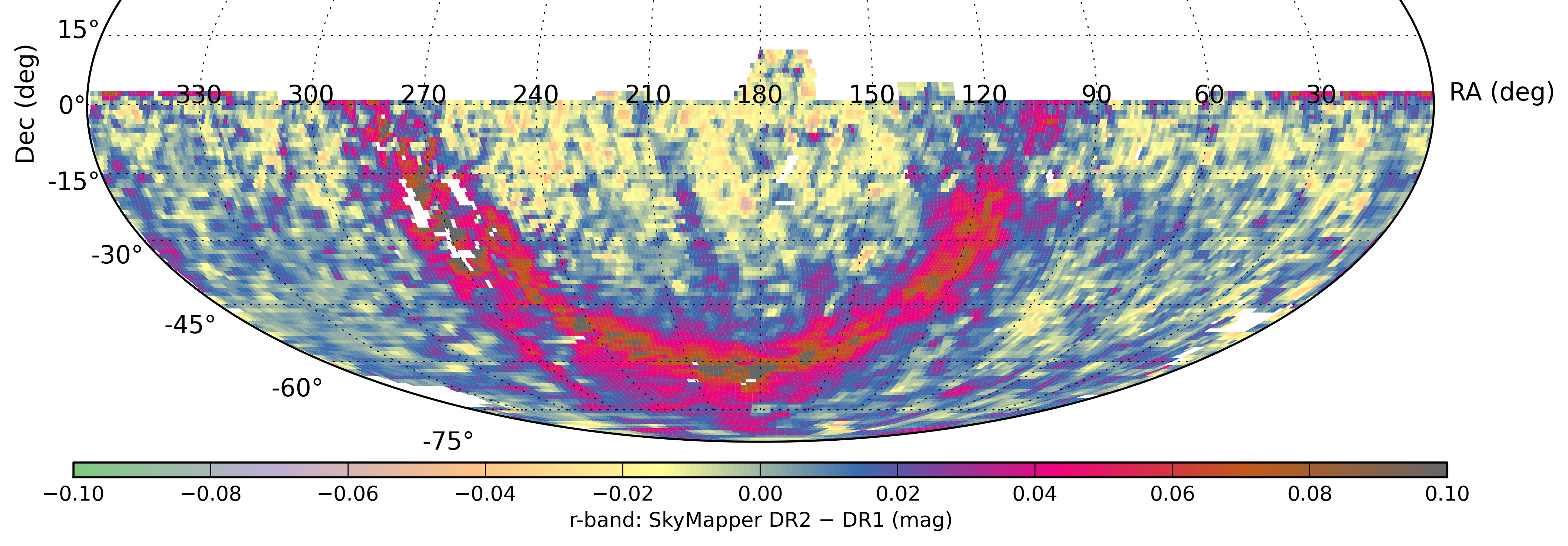}
\includegraphics[width=0.49\textwidth]{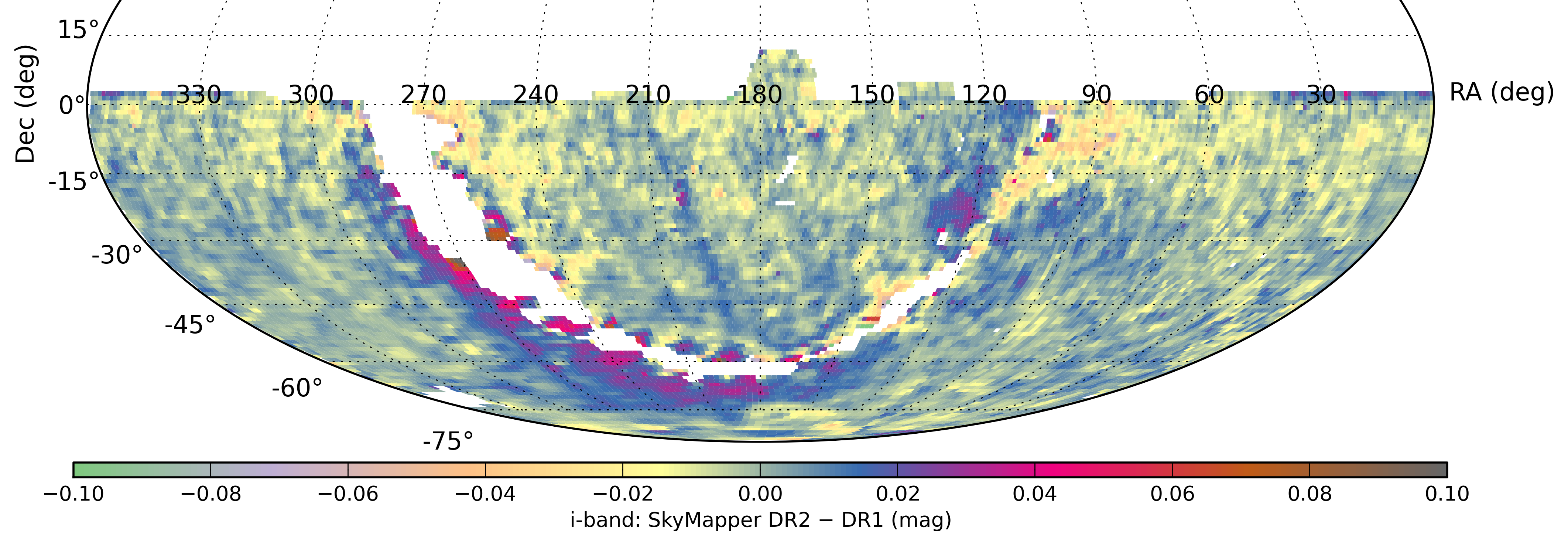}
\includegraphics[width=0.49\textwidth]{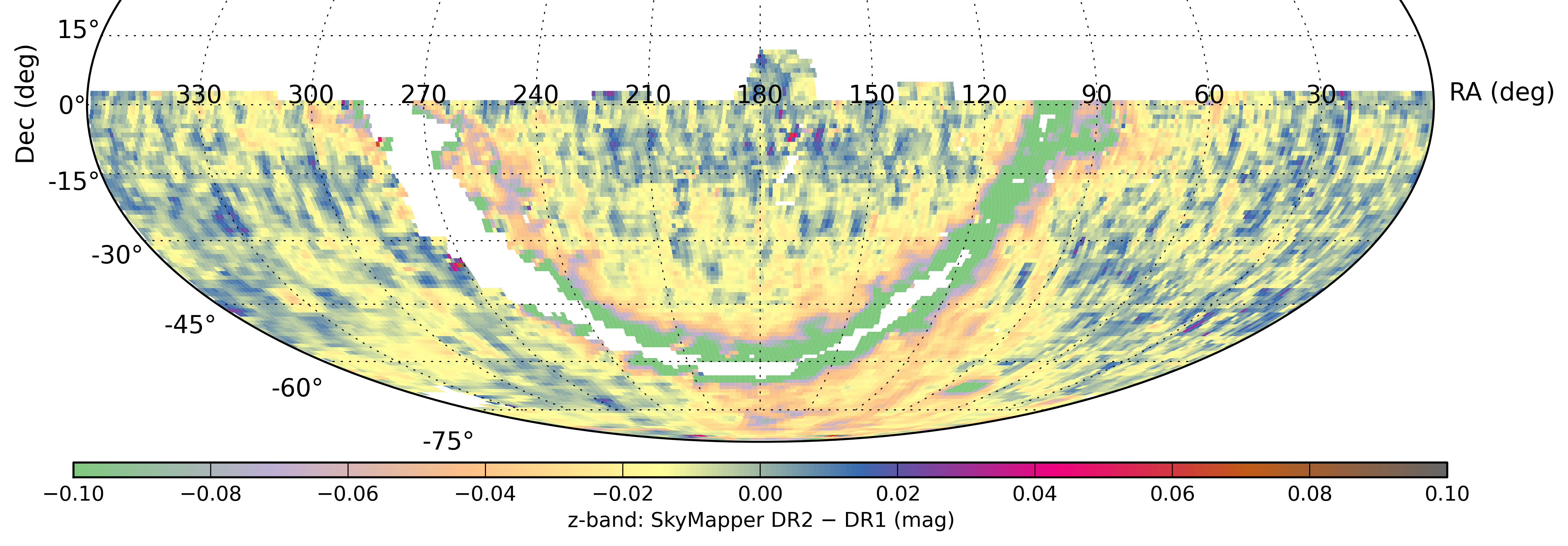}
\caption{Comparison of DR2 photometry with DR1 photometry in all filters: plotted are the median magnitude differences per deg$^2$, restricted to objects with PSF magnitudes brighter than 16. These differences are expected to result primarily from the improved calibration procedure in DR2. }\label{comp_DR2_DR1}
\end{center}
\end{figure*}

\begin{figure*}
\begin{center}
\includegraphics[width=\textwidth]{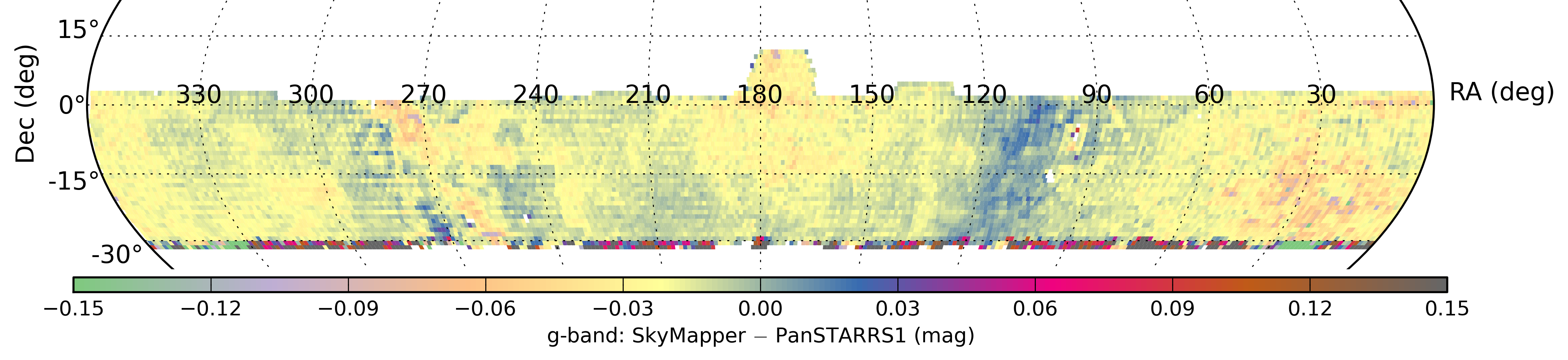}
\includegraphics[width=\textwidth]{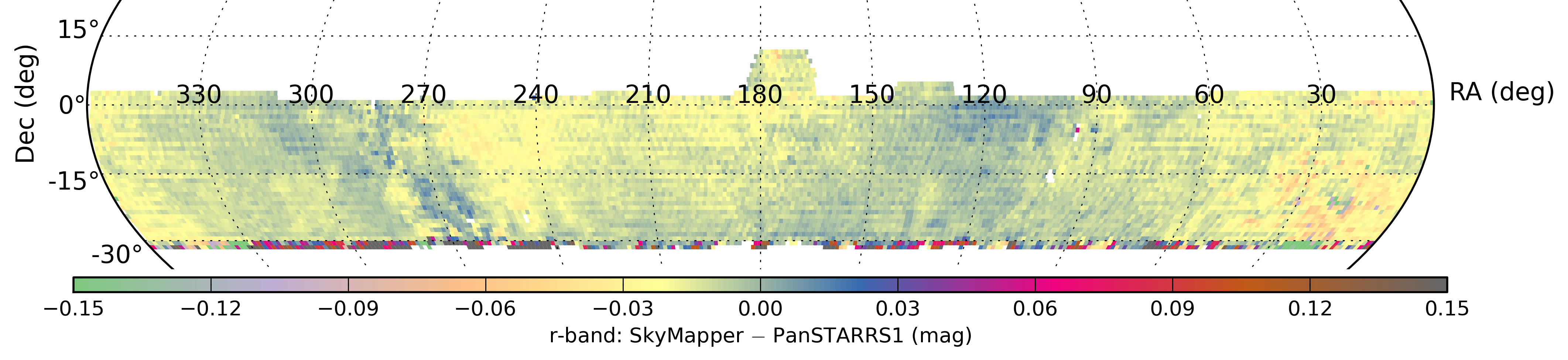}
\includegraphics[width=\textwidth]{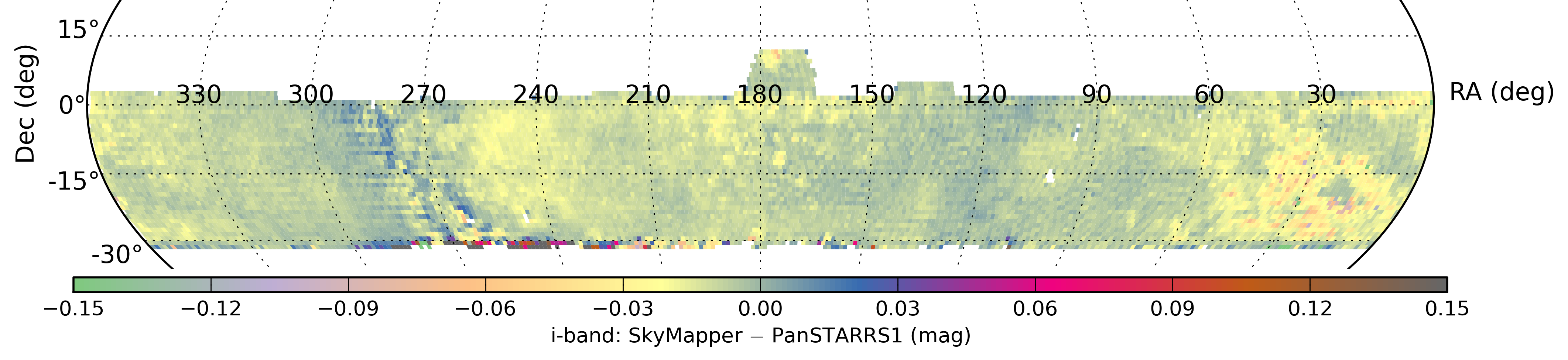}
\includegraphics[width=\textwidth]{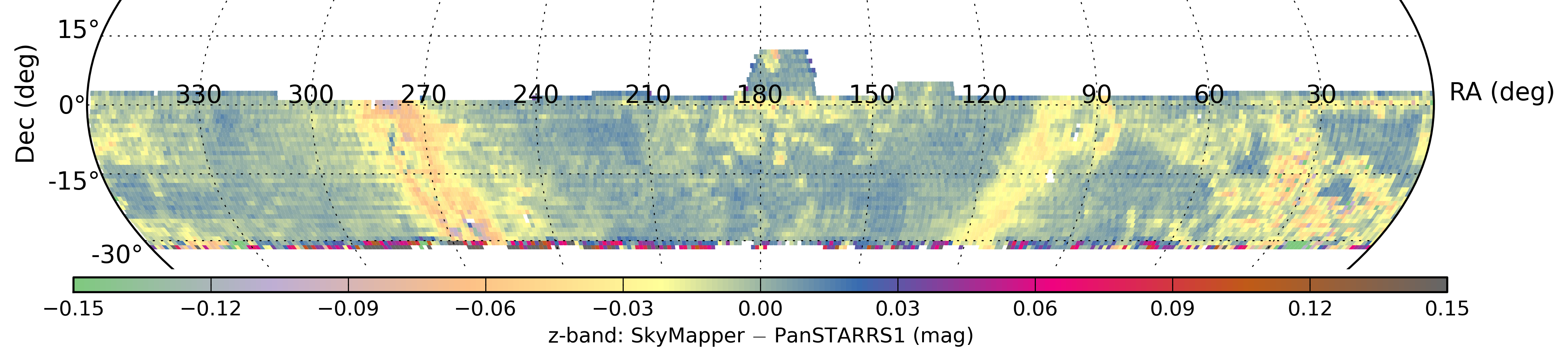}
\includegraphics[width=\textwidth]{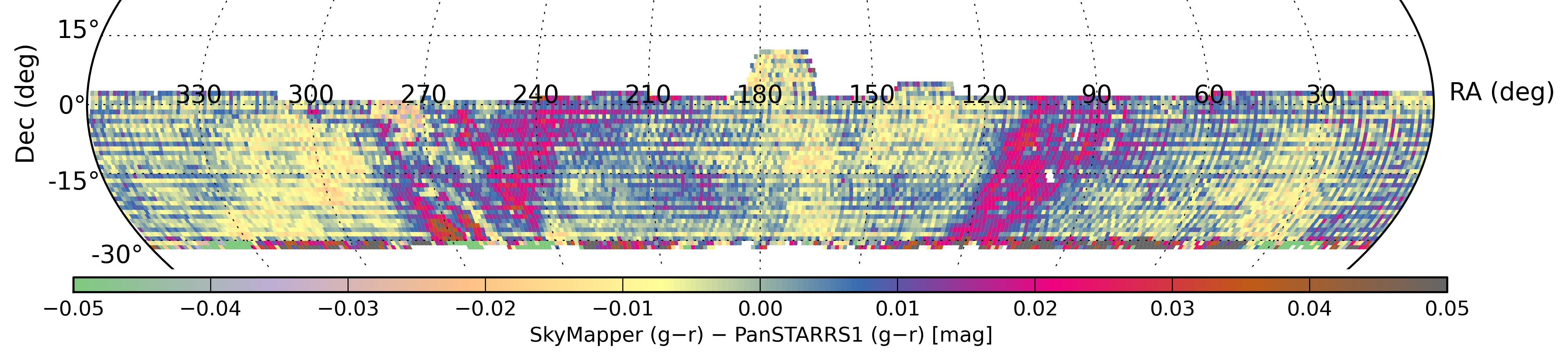}
\caption{Comparison of SkyMapper DR2 photometry with Pan-STARRS1 DR1 photometry in $g,r,i,z$ after applying bandpass transformations by \citet{Tonry18}: plotted are the median magnitude differences per deg$^2$, restricted to objects with PSF magnitudes between 14.5 and 17.5. The large offsets at the Southern edge of the PS1 coverage are where PS1 becomes unreliable. The bottom panel compares the median $g-r$ colours in a high-contrast map and shows that the most extreme differences reach $\pm 0.03$~mag. }\label{comp_SM_PS1}
\end{center}
\end{figure*}

\begin{figure*}
\begin{center}
\includegraphics[width=0.95\textwidth]{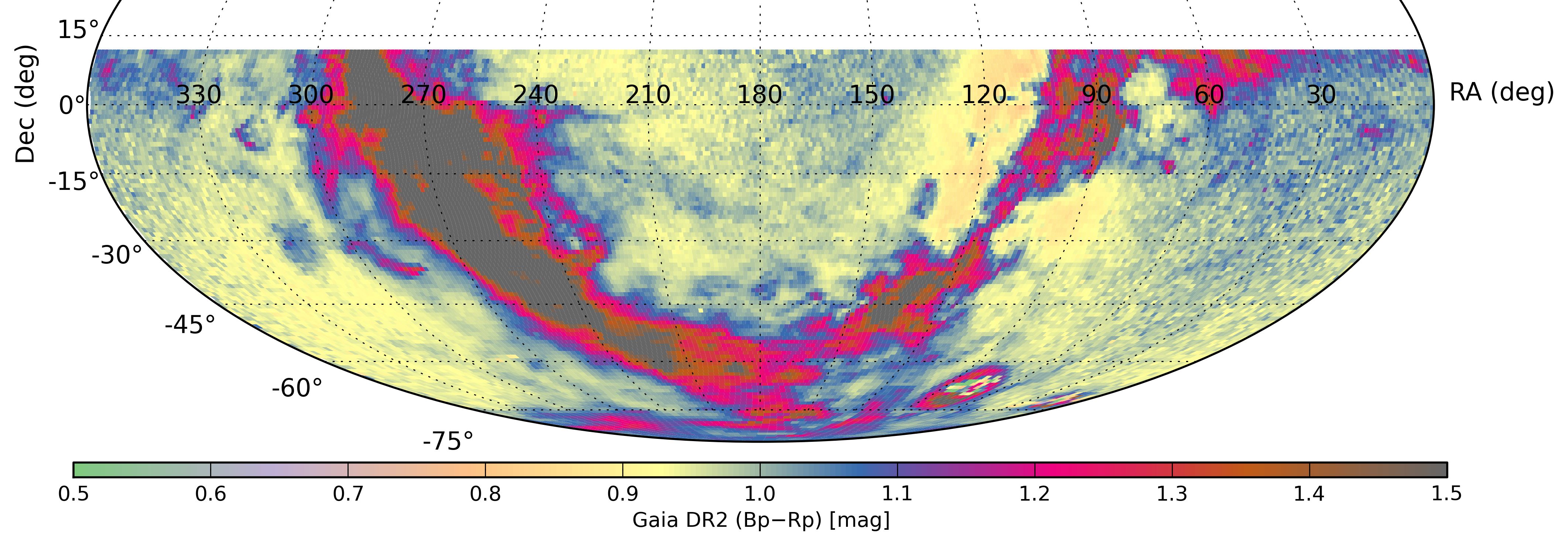}
\includegraphics[width=0.95\textwidth]{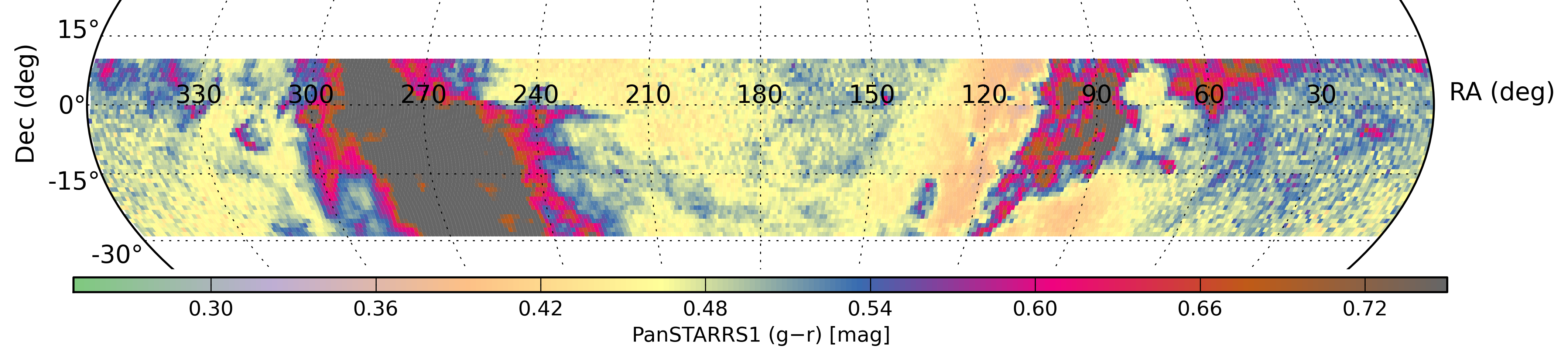}
\includegraphics[width=0.95\textwidth]{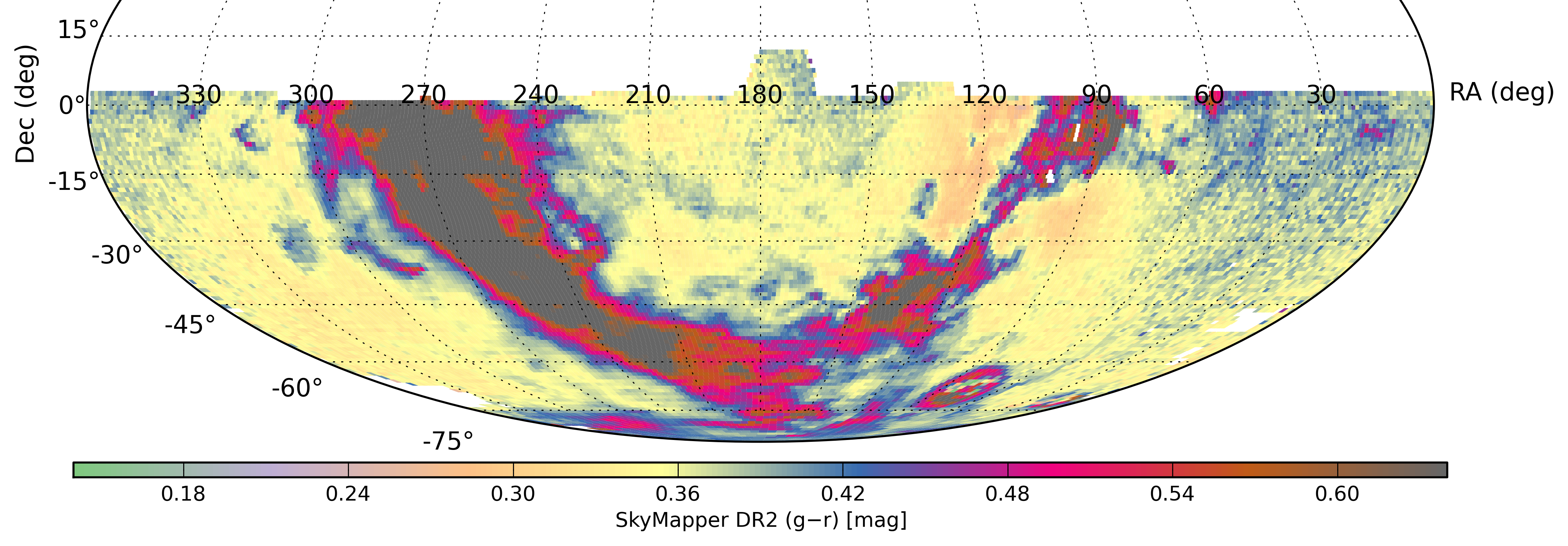}
\includegraphics[width=0.95\textwidth]{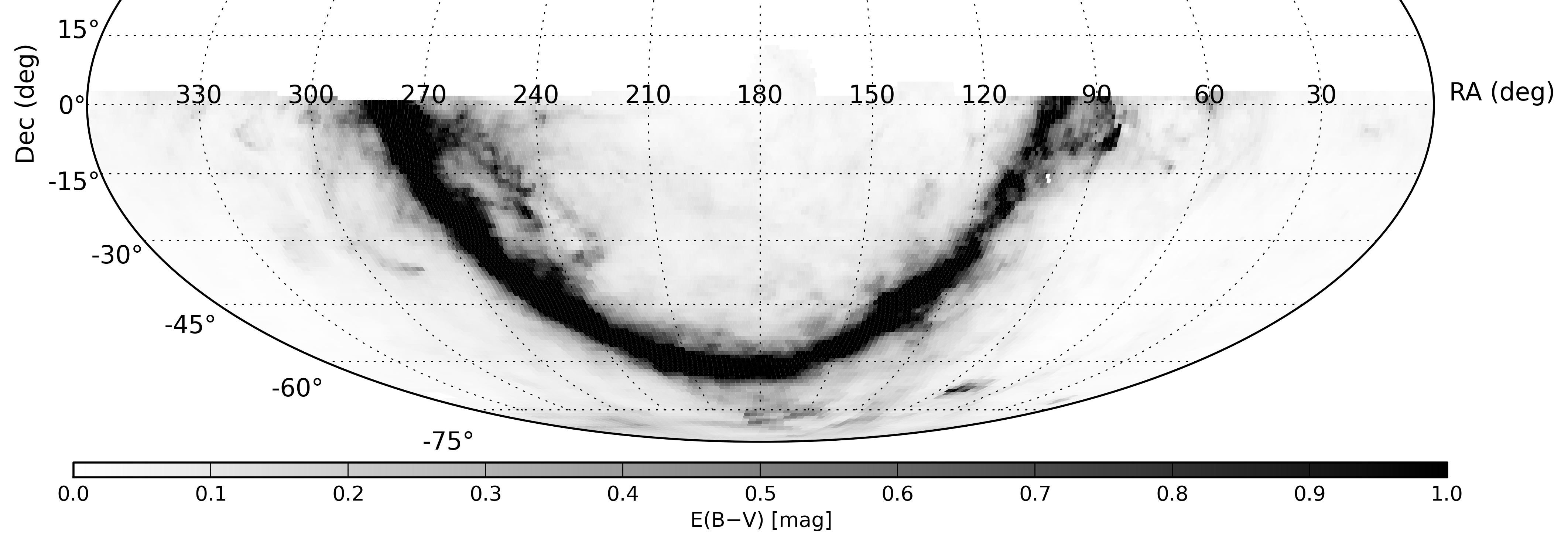}
\caption{Median colour of the stellar population as seen in {\it Gaia} $B_p-R_p$, Pan-STARRS1 $g-r$ and SkyMapper $g-r$: plotted are the median colours per deg$^2$, restricted to objects with magnitudes 14.5 to 17.5. The bottom panel shows the median $E(B-V)$ reddening value from \cite{SFD98}. }\label{skymap_colour}
\end{center}
\end{figure*}

\subsection{Sky coverage}

The mixture of Shallow Survey and Main Survey data and their sky coverage can be seen from Figure~\ref{fig:cov6}, which show, for each filter, the deepest photometric zeropoint for each SkyMapper field. While the zeropoint maps distinguish between the areas with Shallow Survey (light) and Main Survey (dark) images, the FWHM and background levels for each observation serve to further modify the true point source sensitivity across the sky.

The sky coverage of the Main Survey falls into two filter groups owing to the observing strategy: $uvgr$ coverage is driven by the colour sequences that are available for 40\% of the hemisphere, and slight differences in image availability stem from the application of hard quality thresholds to individual images. In contrast, the $iz$ filters cover nearly the whole hemisphere because of twilight observations that are dedicated to this filter pair.

DR2 has tighter limits on image quality than DR1, in particular for zeropoint homogeneity within exposures; this means that users could find data missing in DR2 that was previously part of DR1. Since better data is not always available, some DR1 objects will be absent.

\subsection{Photometric table dataset}\label{sec:photom_table}

The dataset comprises tables with observing information (\texttt{images}, \texttt{ccds}, and \texttt{mosaic}), tables of on-sky photometry, (\texttt{master} and \texttt{photometry}), and copies of external catalogues pre-matched to SkyMapper sources, so that users can quickly generate multi-wavelength table joins.

DR2 provides instantaneous photometry, i.e. measurements made on individual images, as well as averages distilled from repeat measurements, which will be meaningful for non-variable objects. Since objects are only searched on individual images and not on image stacks, the object catalogue is less deep than the dataset would allow in principle.

However, the distilled photometry in the \texttt{master} table reduces errors by combining all individual measurements that are flagged as reliable, and the distilled errors reflect variations beyond photon noise, such as resulting from uneven throughput variations. A planned addition for the next release is searching for objects on deeper co-added frames and performing forced-position photometry, which will create a deeper catalogue and more precise photometry for extended sources.

Most science applications will be served well with data from the \texttt{master} table alone, which contains astrometry and six-filter PSF and Petrosian photometry, as well as flags and cross-link IDs to multi-wavelength tables from external sources. Each row in the \texttt{master} table represents one astrophysically unique object. 

The more detailed \texttt{photometry} table contains one row per unique detection in any SkyMapper image. A single object from the \texttt{master} table may thus appear in several dozen rows in the \texttt{photometry} table, depending on the number of visits to the field and the number of filters in which the object is visible. Objects can be identified or joined to the \texttt{master} table with the OBJECT\_ID column (described further in the next section).

\subsubsection{Master table}

The DR2 \texttt{master} table contains about half a billion (505\,176\,667) unique astrophysical objects. Their photometric measurement can have good flags ({\sc Source Extractor} FLAGS 0 to 3, no other issues) or bad flags ({\sc Source Extractor} FLAGS $>=4$ or other issues). 

Among the half billion objects, about 2.75 million (0.55\%) do not have a single good measurement in any filter, either because they are saturated in all filters or because they are too close to a very bright star and thus flagged to have possibly bad photometry or be bogus detections due to scattered light. 

A further $\sim$82.3 million (16\%) have at least one filter with only one good measurement. The remaining 422\,839\,424 objects (83\%) have more than one good measurement in all filters in which they are detected at all. Filters without any detections do not count towards this consideration. 

The astrometric calibration of DR2 is done as in DR1 and astrometric precision has not changed overall. The median offset between our positions and those in {\it Gaia} DR2 is $0\farcs16$ for all objects and $0\farcs12$ for bright, well-measured objects. In the next release, we plan to switch the astrometric reference frame to {\it Gaia} DR2.

The change in calibration procedure from DR1 to DR2 implies that every object in common between the two releases will have new photometry (and a new OBJECT\_ID).\footnote{Examining changes from DR1 to DR2 is easily accomplished by matching the DR1\_ID column of the DR2 master table to the OBJECT\_ID column of the DR1 master table.} In the $ugri$ filters, the average change among bright stars is less than 1\%, while the average $v$-band magnitude has dimmed by 2.5\%, i.e. it moved by the opposite of what was suggested in \citet{Casagrande19}, and the average $z$-band magnitude has brightened by 1.5\%.

Figure~\ref{comp_DR2_DR1} shows a map of the magnitude differences between our two data releases, which are almost entirely a result of changing the calibration reference; the change from APASS DR9 in our DR1 to {\it Gaia} DR2 in our DR2 imports the all-sky homogeneity of {\it Gaia} into the SkyMapper calibration and removes the substructure we had imported previously from APASS DR9.

When we compare our measured photometry with that predicted from {\it Gaia} for the sample of our zeropoint stars, we find on average no offset in any filter, as expected by design of our calibration procedure. We find, however, RMS dispersions among magnitude differences that range from 1.5\% for $g$ and $r$ filters to 5\% for the $u$-band. This is a result of true physical dispersion, presumably due to the distribution of metallicity values and their effect on measured $u$-band magnitudes, where the predicted magnitudes ignore metallicity and involve just a mean transformation for the overall sky population.

Figure~\ref{comp_SM_PS1} compares our measured photometry with that of Pan-STARRS1 DR1 assuming bandpass transformations by \citet{Tonry18}. We restrict the comparison to the magnitude range $[14.5,17.5]$ for all four filters, where both surveys should be complete and free from saturation effects. The predominant differences are close to the Galactic plane and follow Galactic structure, and so are likely due to the treatment of reddening and/or issues of source density in crowded fields \cite[cf.][]{Tonry18}. The mottled effect in the ($g-r$) map (which is shown at higher contrast than the others) is primarily due to residual flatfielding imperfections in the SkyMapper images.

We also see small-scale structure away from the plane, where star densities should be generally low. Focusing on a small halo region around RA$=180$, we find RMS differences in the map of less than 0.01~mag in $gr$ filters and also in the $g-r$ colour map, although peak-to-peak the $g-r$ colour can range by 0.035~mag in the halo fields at Northern Galactic latitudes. 

We estimate the internal reproducibility of DR2 photometry from repeat measurements of bright stars with good flags and find RMS values in the PSF magnitude of 10~mmag in $u$ and $v$ filter, and 7~mmag in $griz$. 

We also show a map of the median colours of stars as measured by SkyMapper and PS1 ($g-r$) as well as {\it Gaia} ($B_p-R_p$) in Figure~\ref{skymap_colour}. We note that the visible structures are extremely similar. Overall we see the well-known trend with Galactic latitude: halo fields are dominated by red stars seen through little reddening by interstellar dust, while intermediate-latitude fields are populated increasingly by younger and bluer stars; closer to the Galactic plane dust reddening makes the population appear strongly redder, while some special regions on the plane show only unreddened foreground stars, as the most highly reddened stars are invisible. 

\begin{figure}
\begin{center}
\includegraphics[angle=270,width=0.9\columnwidth]{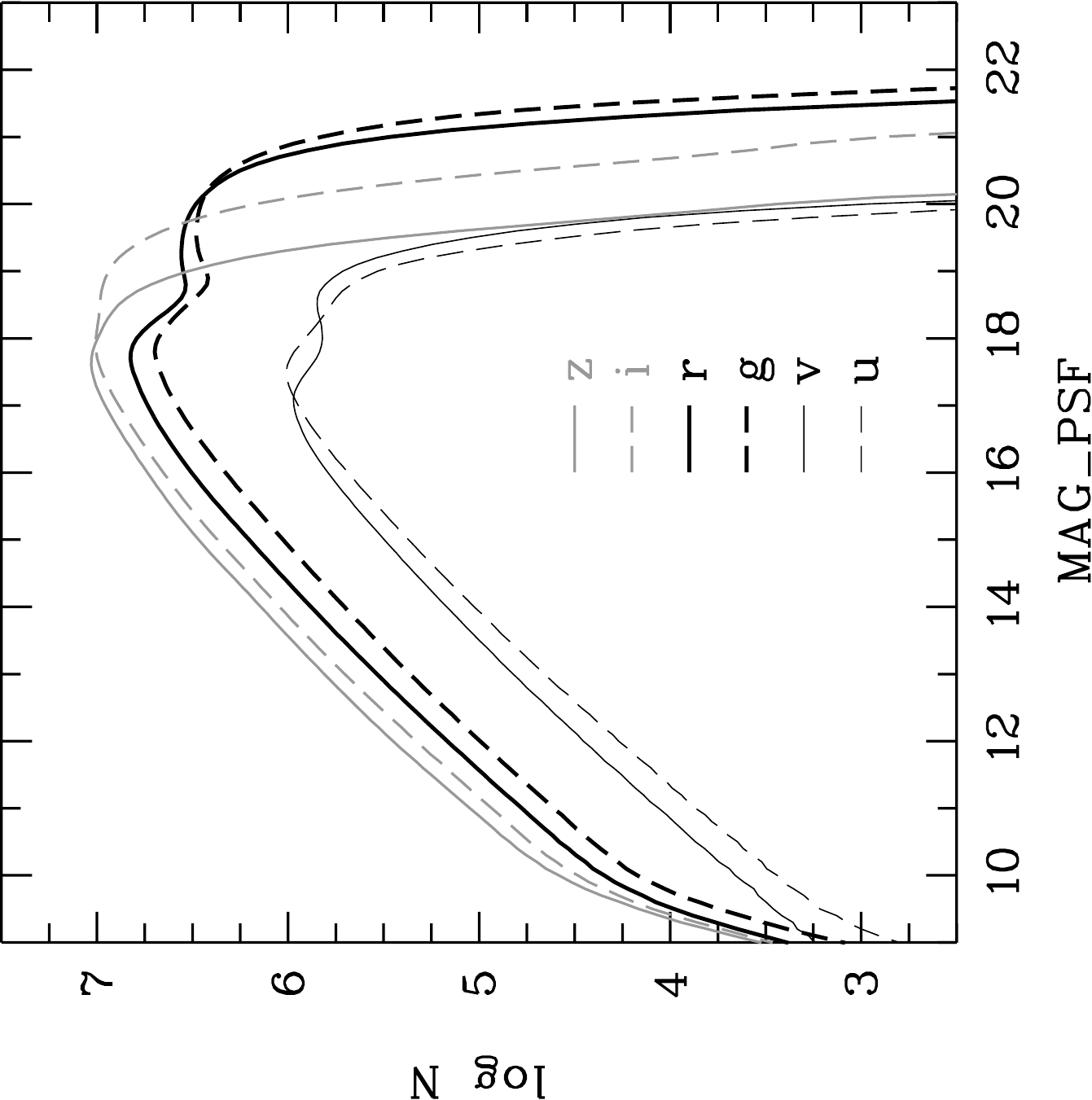}
\caption{Number counts per filter, restricted to objects with $<5$\% error in PSF magnitude. All filters show the sum of two contributions: a large area covered by the Shallow Survey that is complete to nearly 18~mag, and a smaller area covered by the Main Survey that provides deeper data and creates a second maximum.}\label{NC6}
\end{center}
\end{figure}

Finally, Figure~\ref{NC6} shows the number counts from the \texttt{master} table in all six passbands using only objects with PSF magnitude errors of less than 5\%, corresponding to $>20\sigma$-detections. The curve for each filter shows a double-peaked structure as a result of combining two contributions: the Shallow Survey peaks around 17 to 18~mag and covers nearly the whole hemisphere, while the deeper Main Survey peaks around 19 to 20~mag depending on filter but covers a smaller area. In the $i$ and $z$ filters, the double peak is not pronounced because the exposure advantage of the Main Survey translates only into a square-root gain in depth with the sky-limited background in these two filters. Generally, the peaks are softened by a range in sky transparencies and seeing levels mixed in the overall dataset.

\begin{figure*}
\begin{center}
\begin{minipage}[c][10.1cm][t]{.54\textwidth}  
  \vspace*{\fill}
  \centering
  \includegraphics[angle=270,width=10.1cm]{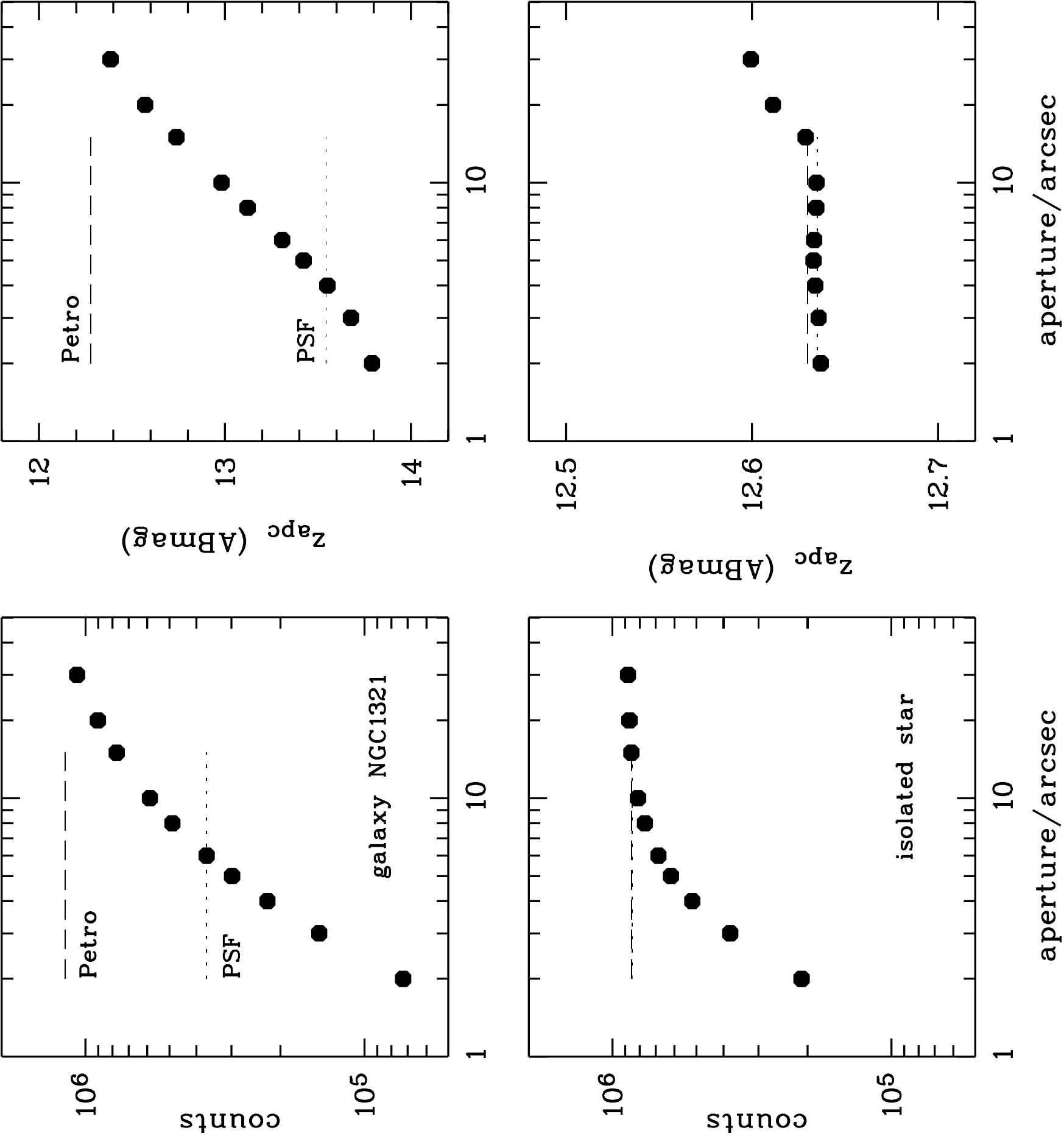}
\end{minipage}  
\raisebox{0.75mm}{
\begin{minipage}[c][8.65cm][t]{.357\textwidth}  
  \centering
  \includegraphics[angle=0,width=3.96cm,frame]{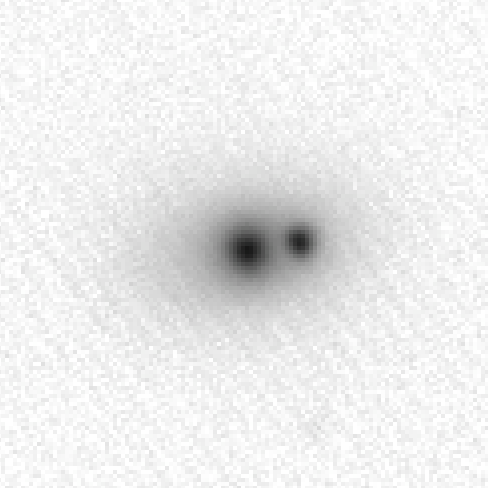}\\  
  \vspace*{\fill}
  \includegraphics[angle=0,width=3.96cm,frame]{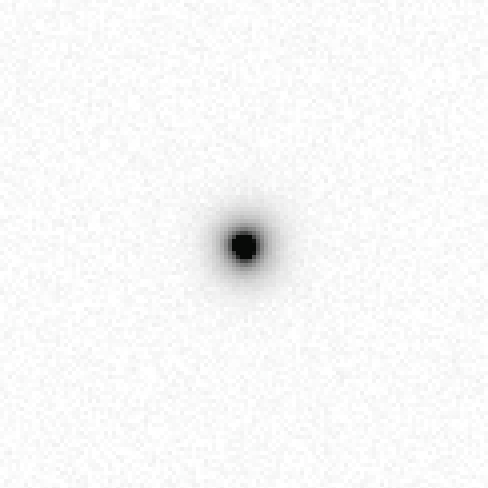}
\end{minipage}
}
\caption{Growth curves of aperture measurements in the \texttt{photometry} table for the double source NGC~1321 (OBJECT\_ID 21671053, a galaxy with a foreground star) and for a star (OBJECT\_ID 21670913) in the same image ($z$-band, FWHM $2\farcs7$). {\it Left:} Count rates measured in apertures with diameters of 2, 3, 4, 5, 6, 8, 10, 15, 20 and 30 arcsec (table columns FLUX\_AP02 to FLUX\_AP30). Overplotted are measured Petrosian and estimated 1D-PSF fluxes, which are identical within 0.5\% for the star, while (unsurprisingly) the PSF flux of the galaxy is just a fraction of the total flux. {\it Centre:} Aperture magnitudes for apertures from $2\arcsec$ to $10\arcsec$ (MAG\_APC02 to MAG\_APC10) are corrected for the growth curve of the expected PSF at the object location assuming that a $15\arcsec$ aperture represents total magnitude. Aperture magnitudes for $15\arcsec$ (MAG\_APR15) to $30\arcsec$ (MAG\_APR30) are as measured. For the star, the various nested apertures predict nearly identical PSF magnitudes after correction, and the scatter among them is $\sim 2$~mmag, however, the uncorrected magnitudes in the larger apertures show that there is an additional $\sim3.5$\% of flux in the wings of the PSF between the $15\arcsec$ and the $30\arcsec$ aperture. {\it Right:} Image cutouts of the two objects, size $60\arcsec \times 60\arcsec$. }
\label{fig:gr_curves}
\end{center}
\end{figure*}

\subsubsection{Photometry table}

The \texttt{photometry} table serves three purposes that go beyond the role of the \texttt{master} table:
\begin{enumerate}
    \item It lists photometry in ten nested apertures ranging from $2\arcsec$ to $30\arcsec$ diameter. Since aperture magnitudes are seeing-dependent, and seeing changes between exposures, we do not distill these quantities.
    \item It lists astrometry and photometry for individual exposures of the object. This can be used to study variability in brightness as well as motions (for examples see Sections \ref{varobjs} and \ref{movobjs}).
    \item It lists photometry that is flagged as potentially (but not necessarily) bad and thus excluded from the results in the \texttt{master} table. Users may find the measurements useful, and the measurements may at times be correct.
\end{enumerate}

Nested-aperture photometry is illustrated in Figure~\ref{fig:gr_curves} for an extended galaxy as well as for a point source. Crucially, we report the aperture data for counts and magnitudes differently: aperture counts are listed as measured and thus represent a raw growth curve; aperture magnitudes, in contrast, are estimates of total magnitude corrected with the local growth curve (MAG\_APCnn, "C" indicating corrected values for an aperture diameter of {\it nn} arcsec) for all apertures smaller than $15\arcsec$. The growth curve is determined by comparing the ratio of the smaller aperture flux to the $15\arcsec$-aperture flux. We do this on a CCD-by-CCD basis, and the fit is allowed to vary linearly in ($x$, $y$) position on the CCD. We choose the $15\arcsec$ aperture as a total-magnitude reference for point source calibration; this magnitude, as well as that from the two larger apertures with diameters of $20\arcsec$ and $30\arcsec$, are listed as measured without correction (MAG\_APRnn, "R" indicating raw values). PSF magnitudes are estimated as a constant-value fit to the inverse variance-weighted sequence of corrected aperture magnitudes. Formally, these fits produce estimates of PSF magnitude that appear precise to a milli-mag-level, while residual calibration errors can be much larger. Figure~\ref{fig:gr_curves} also describes the wings of the SkyMapper PSF and shows that point sources still have a few percent of their flux outside a $15\arcsec$ aperture in median seeing. 

\begin{figure}
\begin{center}
\includegraphics[angle=270,width=\columnwidth]{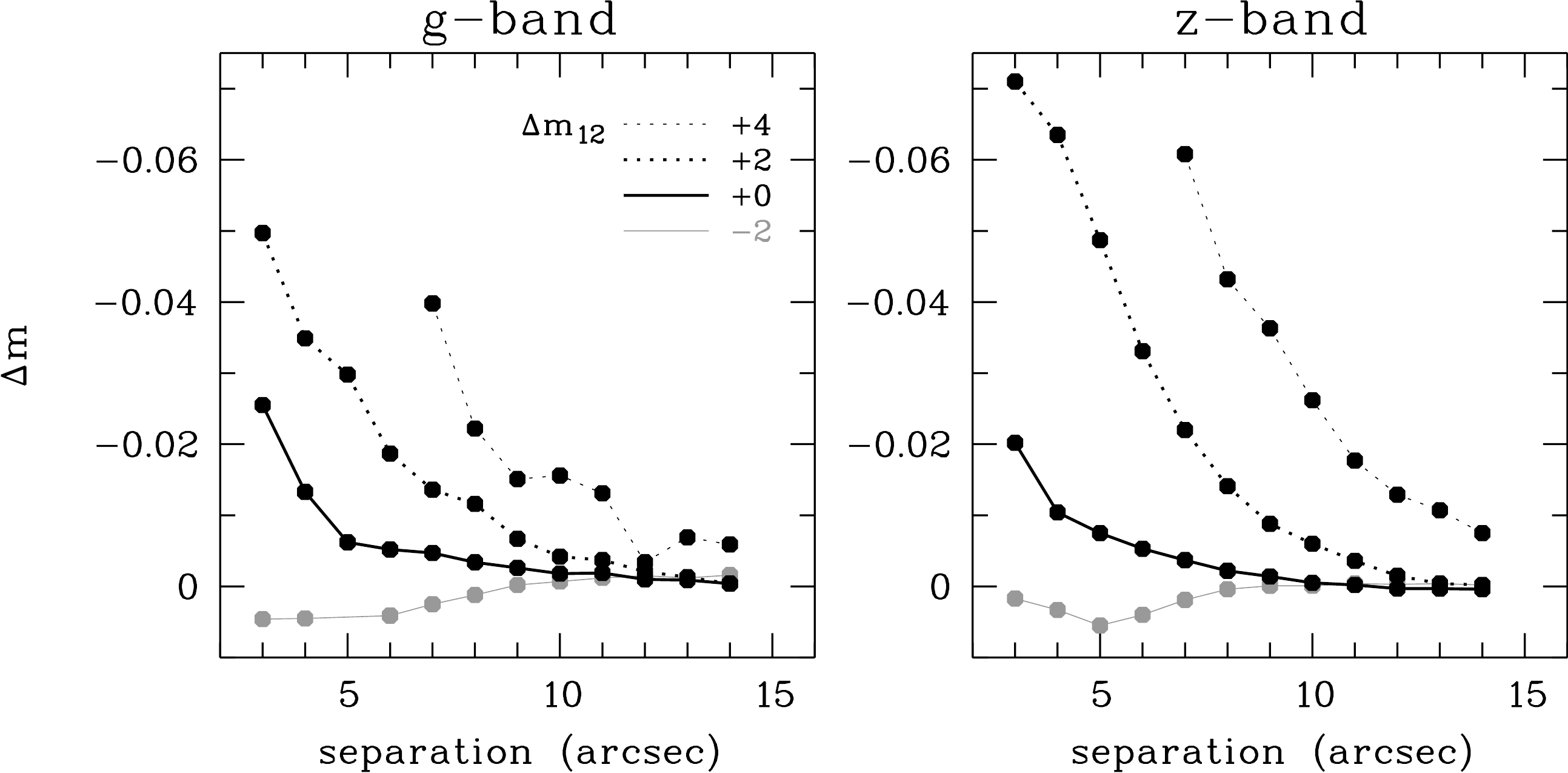}
\caption{Effect of close neighbours on the 1D PSF magnitude of stars: flux from neighbouring stars contributes to the aperture magnitudes of nearby objects and biases their measurement up relative to isolated stars. The effect is $>1$\% for neighbours of equal brightness, when they are closer than $5\arcsec$. Brighter neighbours (positive $\Delta m_{12}$) have an effect already at larger separations, while fainter neighbours can be ignored. }\label{binary_psf}
\end{center}
\end{figure}

\subsection{Limitations of PSF magnitudes}

Our PSF magnitudes are based on one-dimensional (1D) growth curves of point-source light profiles over a $15\arcsec$ diameter, as in DR1. We estimate the precision of these growth curves from the internal reproducibility of the PSF photometry among repeat visits of the same bright objects, and find RMS variations of the PSF flux of 0.7\%. However, our 1D PSF magnitudes are affected by close neighbours as a function of separation $d$ and magnitude difference $\Delta m$, see Figure~\ref{binary_psf}. As a rule of thumb, the PSF magnitudes can be biased brighter by $>1$\% when $d<5\arcsec + 2\arcsec\times \Delta m$. Binary stars of equal brightness only affect each other by $>1$\% if they are closer than $5\arcsec$. Faint sources with bright neighbours, however, can be more strongly affected, even at $15\arcsec$ separation when the neighbour is over 5~mag brighter.

We have expressed this rule of thumb in the column FLAGS\_PSF, where bits 0 to 5 (values 1 to 32) are set when the filters $z$ to $u$ are estimated to be affected by $>1$\%. This rule assumes that both sources are PSF-shaped, and it may be wrong when applied to extended objects. Also, saturated neighbours and those with other bad flags in the \texttt{master} table are assumed to have bad effects out to $15\arcsec$. We did not record neighbours with separations of $>15\arcsec$. If they are sufficiently bright, they might still have an effect, but they would also inhibit the detection of faint neighbours in their PSF wings.

\subsection{Missing table entries}

Users of our data services may encounter situations where image cutouts at the position of a known target show images with a source on them, while the cone search and a full catalogue search reveals an object entry at the position in question, for which the photometry in the relevant filter is missing. How can that be? There are several possible reasons:

\begin{enumerate}
    \item The most common reason will be the bright-star flag that is set for all detections in the vicinity of very bright (-1 to 8~mag) stars. These flags are set for all objects within a radius around the bright stars that depends on the magnitude of the star and filter. Objects in this zone have all measurements flagged within one filter, or some of them if the object is on the edge of the flagging radius. When no unflagged data survives for a filter, the photometry columns in the \texttt{master} table will be empty, and its FLAGS column will reveal the reason. If no filter provides any good measurements, the object's entire spectral energy distribution (SED) will be lost from the \texttt{master} table, but it will be visible in the \texttt{photometry} table.
    \item Another reason may be that all detections of an object in the relevant filter were flagged as bad due to other causes, such as large numbers of bad pixels that are counted in NIMAFLAGS, which have a threshold of 4 for inclusion in the \texttt{master} table, or all detections are saturated (\mbox{FLAGS $=4$} bit being set).
    \item Finally, the parent object identified by the merge algorithm at the location in question might have more than one child in the 
    filter in question, which suppresses the listing of distilled photometry; in this case, the column \{F\}\_NCH (\{F\} being the filter) contains a value $>1$ and the \texttt{photometry} table contains more information.
\end{enumerate}

It is also possible that the \texttt{photometry} table does not reveal entries for an object in an image, which is clearly visible in the image in question. Unless {\sc Source Extractor} overlooked the object, given the parameters we chose, this should only happen when {\sc Source Extractor} extracted two objects in this image while they are considered children of a single parent object for this filter. If all images within one filter show multiple children for what is taken to be a single merged object, all measurements for this object will be missing from the \texttt{photometry} table and hence no distilled summary photometry for this filter can appear in the \texttt{master} table.

\begin{figure}
\begin{center}
\includegraphics[angle=270,width=\columnwidth]{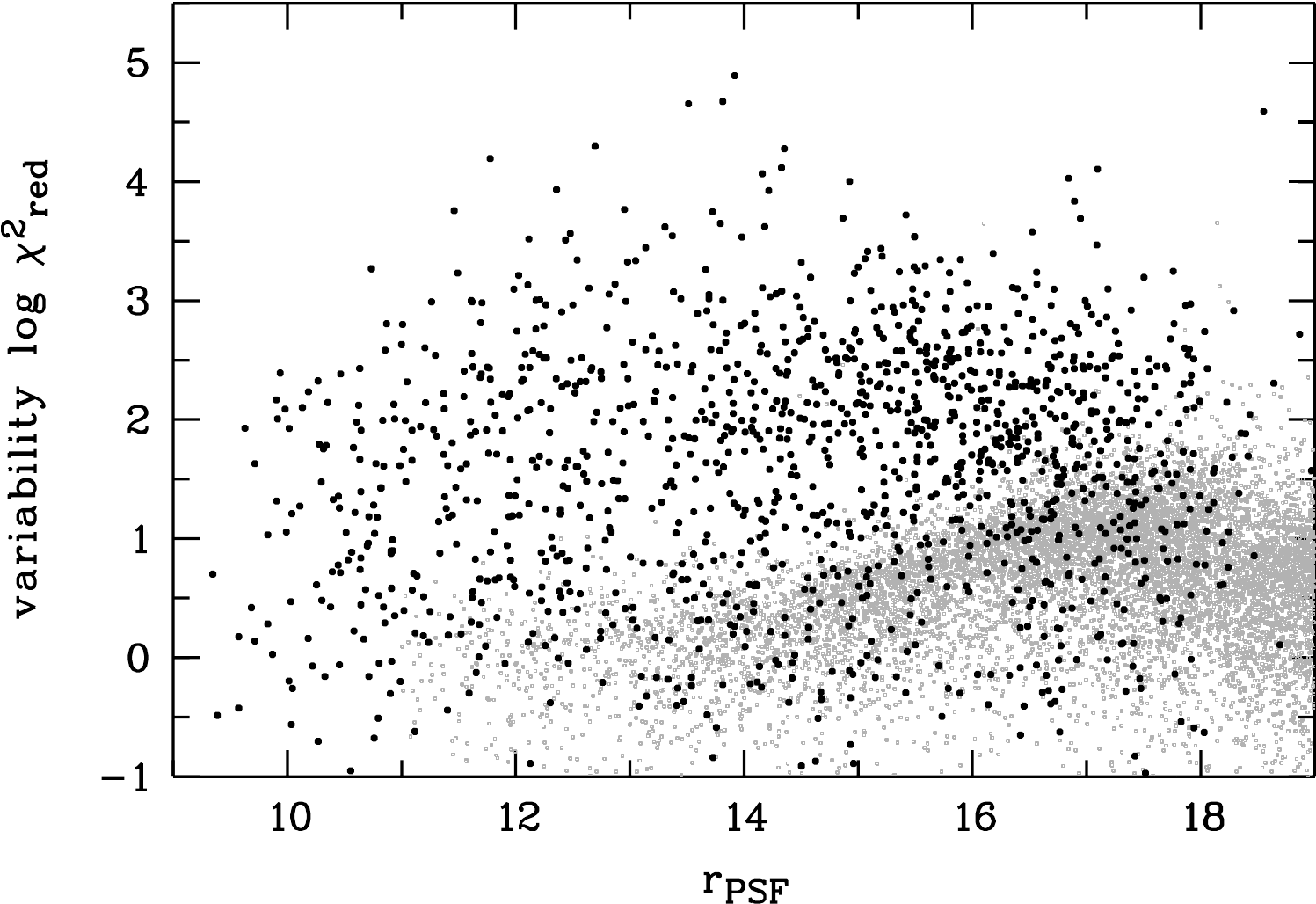}
\caption{Significance of variability detected among DR2 repeat measurements: we compare a random sample (grey) taken from a small sky area with 1\% of the variable source catalogue VSX (large black dots); the plotted variability index is the reduced $\chi^2$ for the set of DR2 photometry being consistent with a constant source, assuming formal flux errors. 
}\label{VSX}
\end{center}
\end{figure}

\begin{figure}
\begin{center}
\includegraphics[angle=270,width=\columnwidth]{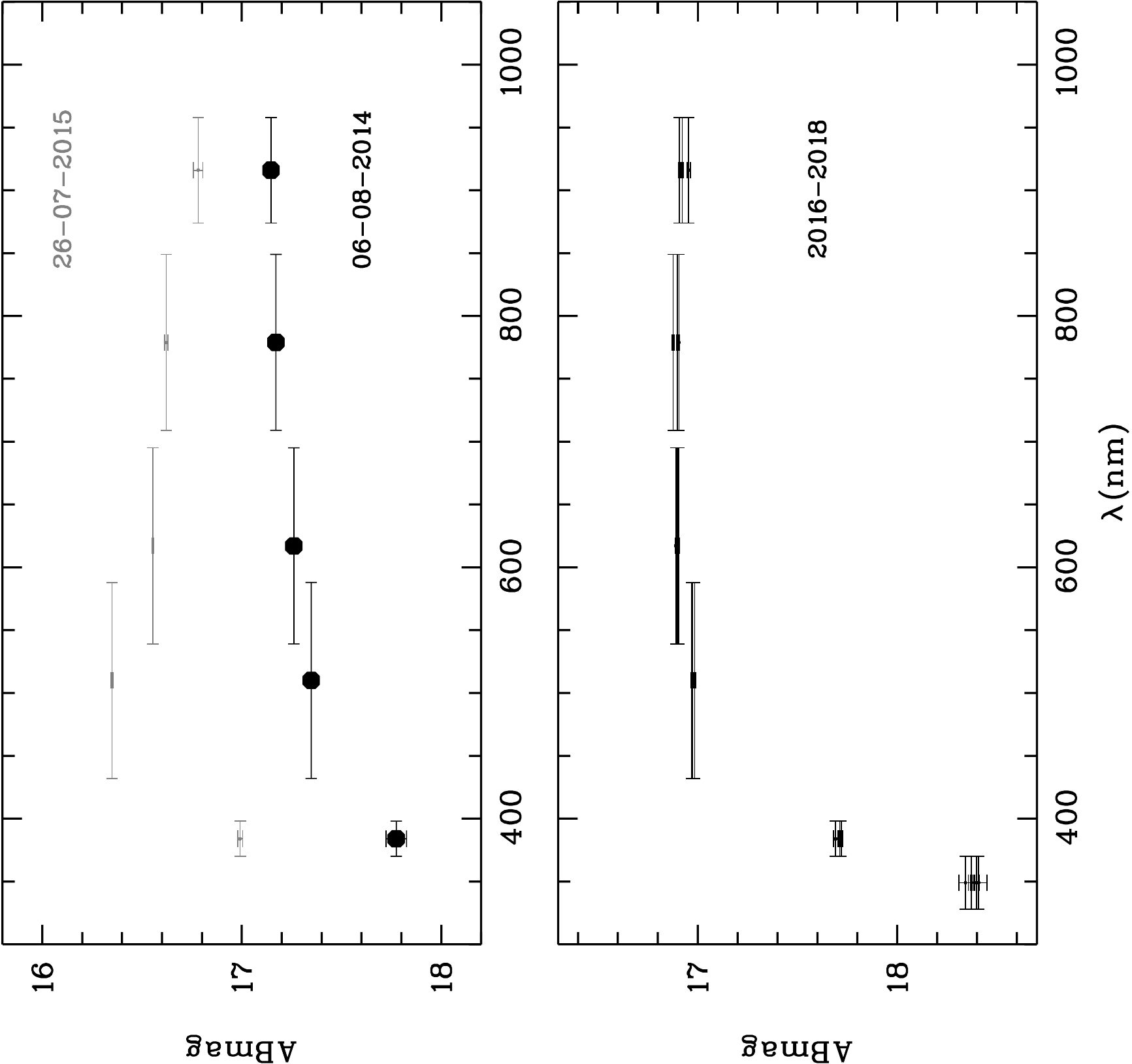}
\caption{Example SEDs of two stars. {\it Top:} a known RR~Lyrae star (OBJECT\_ID 533362) easily recognisable as variable already in the Shallow Survey photometry (two epochs shown). {\it Bottom:} a star (OBJECT\_ID 97057927) seen as not varying even with the better precision of the Main Survey (3 to 4 epochs per filter shown).}\label{varstar}
\end{center}
\end{figure}

\subsection{Variable objects}\label{varobjs}

The observing strategy involves multiple exposures per filter on each sky field, with a range of cadences explained in Section~\ref{sec:strategy}. This allows us to identify variable objects, although the distribution of cadences over filters and survey components means that the selection function will be generally complex and depend on sky position, filter, and object brightness. The \texttt{master} table contains a variability index \{F\}\_RCHI2VAR for each filter \{F\}, which is calculated as the reduced $\chi^2$ among the repeat measurements for the hypothesis of an object being constant in brightness. This measure is driven by both true brightness variations as well as calibration uncertainties and erroneous measurements.

Figure~\ref{VSX} shows a random sample of objects from a small sky area of $2^\circ \times 2^\circ$ in comparison with known variable objects, for which we use a random 1\% subset of the objects from the AAVSO International Variable Star Index \citep[VSX;][]{VSX06,VSXcat}. Most of the known variable stars are clearly distinguished by a high variability index. New variables can be identified using suitable thresholds, but any analysis of their SEDs for classification purposes would ideally consider the individual detections in the \texttt{photometry} table.

The 6-filter SED of variable objects in the \texttt{master} table can be entirely wrong, because filters are observed at different times and clipped for outliers from a median estimate in the distill process. In Figure~\ref{varstar} we compare a known variable star of type RR~Lyrae to a random star seen as not varying. The SED of the variable star is plotted from the \texttt{photometry} table for two epochs with Shallow Survey photometry that captures all six filters within a few minutes, while the non-varying star is represented with three to four (depending on filter) epochs. Error bars on magnitudes are mostly too small to be visible, and horizontal error bars represent the filter FWHM.

\begin{figure}
\begin{center}
\includegraphics[width=\columnwidth]{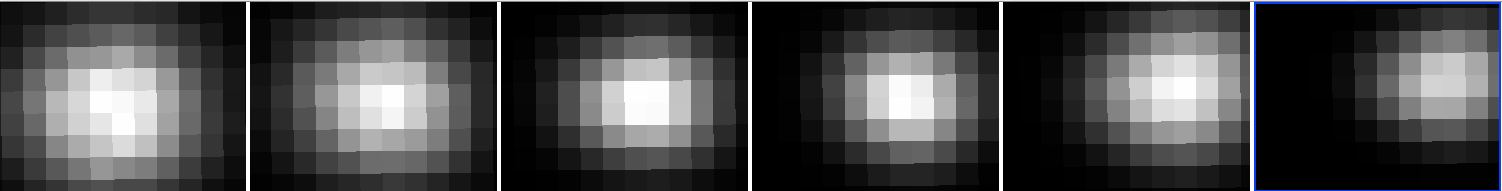}
\caption{Time/filter sequence of the Shallow Survey visit from 18 June 2014 to the $g\approx 15$ M~star with object ID 282264432. Pixel scale is the usual $\sim 0\farcs5$. In these images the dominant source here is actually the $V=10.6$~mag asteroid (230) Athamantis, which was moving with $\sim 0\farcs5/$min. From left to right the filters run along the usual Shallow Survey sequence $uvgriz$, and the middle of the first and last exposure are separated by 177~sec. 
}\label{Mflare_blend}
\end{center}
\end{figure}

\begin{table*}
\caption{Example of a moving object: multiple appearances of the dwarf planet Pluto in the \texttt{master} table.}
\label{Pluto}
\centering
\begin{tabular}{lcccccccccc}
\hline\hline
 Object ID & RA (J2000) & Dec (J2000) & $u_{\rm PSF}$ & $v_{\rm PSF}$ & $g_{\rm PSF}$ & $r_{\rm PSF}$ & $i_{\rm PSF}$ & $z_{\rm PSF}$ & Date \\ 
\hline
245765990 & 282.76775 & -20.31218 & 16.610 & 16.206 &        &        & 13.782 & 13.778 & 08 Jul 2014\\ 
245561233 & 282.21689 & -20.40180 &        &        & 14.593 & 14.119 & 13.871 & 13.848 & 30 Jul 2014\\ 
258148035 & 284.68803 & -20.77926 & 16.514 & 16.106 & 14.523 & 14.049 & 13.801 & 13.788 & 19 Jul 2015\\ 
259666945 & 290.56376 & -21.19184 &        &        &        &        & 14.073 & 14.015 & 07 Apr 2017\\ 
259666968 & 290.58678 & -21.19338 & 16.761 & 16.214 & 14.662 & 14.226 & 13.934 & 13.921 & 11 Apr 2017\\ 
\hline
\end{tabular}
\end{table*}

\subsection{Moving and transient objects}\label{movobjs}

Some objects in the \texttt{master} table have only one detection, although that region has been visited multiple times. This includes asteroids and dwarf planets that will be seen at different times in different sky locations, as well as transients that are stable in location but have such high variability amplitudes that they exceed the detection threshold of our imaging only on one occasion. The latter category includes flares on M~dwarf stars that are too faint in quiescent state, as well as novae and supernovae. 

Here, we note two examples: first, the 14th magnitude dwarf planet Pluto appears five times in the \texttt{master} table between July 2014 and April 2017 (see Table~\ref{Pluto}). A second, noteworthy, case was identified while selecting M~dwarf flares from DR1 (Chang et al., in preparation): an apparent super-flare was seen in an M~giant (OBJECT\_ID 282264432 in DR2) on 18 June 2014, that seemed to brighten the star to $g\approx 10.7$~mag relative to the other detections around 15~mag. This event turned out to be a chance blend of the star with the $V=10.6$~mag Inner Main Belt asteroid (230) Athamantis, which was moving about half an arcsecond per minute\footnote{We identified the nature of the asteroid using the website of the VO Solar System portal at \url{http://vo.imcce.fr/webservices/skybot/skybotconesearch\_query.php}}. Figure~\ref{Mflare_blend} shows the time/filter sequence of the Shallow Survey visit to the target location on 18 June 2014 around 16:40~UT, where the middle of the first and last exposures are separated by three minutes. Due to the motion, the $u$ and $v$ centroids differ from the star's own location such that blended $uv$ magnitudes appear in a separate entry in the \texttt{master} table with OBJECT\_ID 282264433.

\subsection{Images and CCDs tables}\label{sec:im_table}

The \texttt{images} table lists dates, positions, exposure times, and quality indicators of each individual telescope exposure. It can be used to differentiate between Shallow Survey and Main Survey images, based on the column IMAGE\_TYPE, which is 'fs' for the Shallow Survey\footnote{The 'fs' label originates from when the Shallow Survey was known as the "Five-Second Survey", based on its shortest exposure times.} and 'ms' for the Main Survey, or based on EXP\_TIME, which is 100~sec for the Main Survey but shorter for the Shallow Survey. Images are identified by a unique IMAGE\_ID, which encodes roughly the UT date and time of the shutter opening for the exposure in the format YYYYMMDDhhmmss. While the value is rounded to 1~sec, it is not actually the time stamp for the exposure start: it is on average 2~sec earlier than the exposure start and extreme values in DR2 range from $-11$~sec to $+9$~sec. (The DATE column contains the actual MJD at the start of the exposure.) Image IDs are listed for every object detection in the \texttt{photometry} table.

The \texttt{ccds} table contains every valid CCD of every exposure, and thus nearly 32 times as many rows as the images table. Occasionally, CCDs will be missing for some exposures, either because an adequate World Coordinate System solution could not be determined or because the readout amplifiers failed. The COVERAGE column defines the polygon of a CCD's footprint, and could be used to check whether a known target location has fallen on a CCD that is part of the data release. This information can be used to investigate non-detections of known targets. An alternative approach for a small number of targets is to visually inspect small images served by the cutout service, which will return all release images at the target location.

\section{DR2 Data Access}\label{sec:access}

Access to DR2 catalogues and images is provided through the SkyMapper node of the All-Sky Virtual Observatory (ASVO), a framework that builds upon the standards of the International Virtual Observatory Alliance (IVOA). Direct access to the tables and image cutouts is possible using the Table Access and Simple Image Access Protocols (TAP and SIAP), and a Simple Cone Search (SCS) functionality enables position-based object searches. The SkyMapper website\footnote{http://skymapper.anu.edu.au} augments those services with an 'object viewer' summarising the key information for each object in the \texttt{master} table, a 'spectrum viewer' to display associated spectroscopic data (initially with the 6dFGS spectra, and expanding to include other datasets), and a User Forum for obtaining help and posting tips for others.

The changes from the DR1 catalogue structure described by \citet{Wolf18a} include:
\begin{enumerate}
    \item The \texttt{master} table adds a measure of the local source density (the number of \texttt{master} table sources within 15~arcsec), a per-filter count of the number of observations sigma-clipped from the final estimate of the PSF magnitude, a per-filter reduced $\chi^2$ statistic for the source being non-variable, and a flag indicating for each of the six filters whether the PSF magnitudes are likely affected by close neighbours at more than a 1\% level. 
    \item The \texttt{master} table drops the shape measurements provided in DR1, as their robust estimates are difficult with the mix of exposure times and seeing conditions, and drops the visit counts in each filter, since the different photometric depths of the Main Survey and Shallow Survey render such measures difficult to construct in a meaningful manner.
    \item The cross-matching to external catalogues now includes SkyMapper DR1, 2MASS, AllWISE, ATLAS Refcat2, {\it GALEX} GUVcat, {\it Gaia} DR2, PS1 DR1, and UCAC4. The column containing the second-closest 2MASS source has been dropped from the \texttt{master} table in favour of the second-closest {\it Gaia} DR2 source. The 2MASS cross-matching has also been restricted to the PSC, allowing the column specifying the catalogue to be omitted. Instead, it is {\it Gaia} DR2 for which we record the two closest matches, as {\it Gaia}'s superb resolution can flag sources blended in SkyMapper.
    \item A number of smaller external catalogues have been updated to include their closest DR2 cross-matches: 2dFGRS, 2dFLenS, 2MRS, 2QZ/6QZ, 6dFGS, GALAH, HES QSO, Milliquas, and VSX.
    \item Since DR2 includes both Main Survey and Shallow Survey images, the name of the photometric table has been generalised to \texttt{photometry}.
    \item The \texttt{photometry} table adds two columns: a boolean indicator of whether a particular measurement contributed to the clipped mean magnitude in the \texttt{master} table, and a $\chi^2$ value for that observation relative to the clipped mean magnitude.
\end{enumerate}

\section{Future Data Releases}\label{sec:future}

The next data release will contain more images and better sky coverage for the Main Survey, as well as co-added sky tiles, where we homogenise the PSFs of images and then re-register and co-add them within filters. The co-added tiles will combine data from both the Main Survey and the Shallow Survey to capture the depth of the images while using shallow data on pixels where deep images are saturated, providing images with wide dynamic range and consistent PSFs across the filters.

In the future, we will also do source-finding on co-added frames, which will give us deeper detections than now; presently, our completeness is limited by detections in individual images even though the distilled photometry has relatively low errors due to the combination of all good detections into distilled magnitudes. Forced-position photometry then becomes possible as well. 

Irrespective of co-added frames we aim to include PSF magnitudes that are based on two-dimensional PSF-fitting instead of 1D growth curves, and are thus more reliable in crowded fields or generally for objects with close neighbours.

We plan processing enhancements such as astrometry tied to {\it Gaia} DR2 as a reference frame and better fitting of electronic interference and CCD bias, especially in areas covered by large galaxies and extended nebulae, where at present the bias is incorrect, causing excess noise and oversubtraction of the background. This is relevant for the creation of high-quality co-added images of galaxies and accurate SEDs of large galaxies.

Finally, we plan to update the photometric calibration: presently the calibration uses a transformation from {\it Gaia} to SkyMapper that assumes a single-parameter family of stars. Especially in the $u$ and $v$ filters, however, colours depend explicitly on metallicity and stellar population gradients across the sky can cause artificial zeropoint drifts. These may be removed by an iterative approach that fits the metallicity from the photometry \citep[e.g.][]{Casagrande19,Huang19} and thus refines the zeropoint table. 

At smaller spatial scale we will tidy up the calibration by improving our flatfields, which currently have issues at the perimeter of the mosaic and can be wrong by up to 5\% in the very corners. However, overall the internal reproducibility of flux measurements of bright stars shows already an RMS scatter of only 1\% in the $uv$ filters and 0.7\% in $griz$.

The main advantages of using DR2 over DR1 are the improved zeropoint calibration and the availability of deeper Main Survey images on part of the hemisphere, both of which enhance the utility of the SkyMapper Southern Survey for extragalactic research and for Galactic archaelogy studies. We note in particular that the $i$ and $z$ filters cover $\sim 90$\% of the hemisphere already, although not yet at the final number of visits and depth.

\begin{acknowledgements}
The national facility capability for SkyMapper has been funded through ARC LIEF grant LE130100104 from the Australian Research Council, awarded to the University of Sydney, the Australian National University, Swinburne University of Technology, the University of Queensland, the University of Western Australia, the University of Melbourne, Curtin University of Technology, Monash University and the Australian Astronomical Observatory. Parts of this project were conducted by the Australian Research Council Centre of Excellence for All-sky Astrophysics (CAASTRO), through project number CE110001020. BPS acknowledges support from the ARC Laureate Fellowship FL0992131. LS expresses thanks for support by the China Postdoctoral Science Foundation (2018M641067).
We acknowledge support from the ARC Discovery Projects program, most recently through DP150103294 (GDC, MSB, DM). DM also acknowledges support from an ARC Future Fellowship FT160100206. Development and support for the SkyMapper node of the All-Sky Virtual Observatory (ASVO) has been funded in part by Astronomy Australia Limited (AAL) and the Australian Government through the Commonwealth's Education Investment Fund (EIF) and National Collaborative Research Infrastructure Strategy (NCRIS), particularly the National eResearch Collaboration Tools and Resources (NeCTAR) and the Australian National Data Service Projects (ANDS). The National Computational Infrastructure (NCI), which is supported by the Australian Government, has contributed resources and services to this project and hosts the SkyMapper node of the ASVO. We acknowledge the Gamilaroi people as the traditional owners of the land on which the SkyMapper Telescope stands.

We thank the referee for comments that improved the quality of the manuscript. We thank John Tonry and Luca Casagrande for extensive discussions that led to the improved photometric zeropoint calibration of DR2. We thank Marc White, Jon Smillie, Sean Pringle, and Chris Allen for their support of the SkyMapper node of the ASVO. 
This work has made use of data from the European Space Agency (ESA) mission {\it Gaia}\footnote{https://www.cosmos.esa.int/gaia}, processed by the {\it Gaia} Data Processing and Analysis Consortium\footnote{https://www.cosmos.esa.int/web/gaia/dpac/consortium} (DPAC). Funding for the DPAC has been provided by national institutions, in particular the institutions participating in the {\it Gaia} Multilateral Agreement. This publication makes use of data products from the Two Micron All Sky Survey, which is a joint project of the University of Massachusetts and the Infrared Processing and Analysis Center/California Institute of Technology, funded by the National Aeronautics and Space Administration and the National Science Foundation. The Pan-STARRS1 Surveys (PS1) and the PS1 public science archive have been made possible through contributions by the Institute for Astronomy, the University of Hawaii, the Pan-STARRS Project Office, the Max-Planck Society and its participating institutes, the Max Planck Institute for Astronomy, Heidelberg and the Max Planck Institute for Extraterrestrial Physics, Garching, The Johns Hopkins University, Durham University, the University of Edinburgh, the Queen's University Belfast, the Harvard-Smithsonian Center for Astrophysics, the Las Cumbres Observatory Global Telescope Network Incorporated, the National Central University of Taiwan, the Space Telescope Science Institute, the National Aeronautics and Space Administration under Grant No. NNX08AR22G issued through the Planetary Science Division of the NASA Science Mission Directorate, the National Science Foundation Grant No. AST-1238877, the University of Maryland, Eotvos Lorand University (ELTE), the Los Alamos National Laboratory, and the Gordon and Betty Moore Foundation. This research made use of {\sc Astropy}, a community-developed core {\sc Python} package for Astronomy (Astropy Collaboration, 2013). 

\end{acknowledgements}


\bibliographystyle{pasa-mnras}

\end{document}